\documentclass[aps,prb,superscriptaddress,amsmath,amssymb,showpacs,floatfix,reprint,longbibliography]{revtex4-2}

\usepackage{graphicx}
\usepackage{float}

\usepackage{multirow}    
    
\usepackage{color}
\usepackage[colorlinks=true, urlcolor=blue, linkcolor=blue, citecolor=blue, pdftex]{hyperref}

\begin{document}

\title{The dynamical structure factor of the SU(3) Heisenberg chain : The  variational Monte Carlo approach}

\author{D\'aniel V\"or\"os}
\affiliation{Department of Physics, Budapest University of Technology and Economics, 1111 Budapest, Hungary}
\affiliation{Institute for Solid State Physics and Optics, Wigner Research Centre for Physics, H-1525 Budapest, P.O. Box 49, Hungary}
\author{Karlo Penc}
\affiliation{Institute for Solid State Physics and Optics, Wigner Research Centre for Physics, H-1525 Budapest, P.O. Box 49, Hungary}

\date{\today}

\begin{abstract}
We compute the dynamical spin structure factor $S(k,\omega)$ of the SU(3) Heisenberg chain variationally using a truncated Hilbert space spanned by the Gutzwiller projected particle-hole excitations of the  Fermi sea,  introduced in [B. Dalla Piazza {\it et al.}, Nature Physics {\bf 11}, 62 (2015)], with a modified importance sampling.
We check the reliability of the method by comparing the $S(k,\omega)$  to exact diagonalization results for 18 sites and to the two-soliton continuum of the Bethe Ansatz for 72 sites. We get an excellent agreement in both cases. Detailed analysis of the finite-size effects shows that the method captures the critical Wess-Zumino-Witten SU(3)$_1$ behavior and reproduces the correct exponent, with the exception of the size dependence of the weight of the bottom of the conformal tower. We also calculate the single-mode approximation for the SU($N$) Heisenberg model and determine the velocity of excitations. Finally, we apply the method to the SU(3) Haldane-Shastry model and find that the variational method gives the exact wave function for the lowest excitation at $k=\pm 2\pi/3$.
\end{abstract}

%\pacs{75.10.-b, 75.10.Jm, 75.10.Pq, 75.30.Kz, 75.40.Mg}

\maketitle

%%%%%%%%%%%%%%%%%%%%%%%%%%%%%%%%%%%%%%%%%%%%%%%%%%%%%%%
\section{Introduction}
\label{sec:Introduction}
%%%%%%%%%%%%%%%%%%%%%%%%%%%%%%%%%%%%%%%%%%%%%%%%%%%%%%%

% Motivation
One of the most important quantities describing the state of a magnetic material is the dynamical structure factor 
\begin{equation}
  S^{\alpha\alpha} (\mathbf{q},\omega) 
  \propto \sum_{\mathbf{R}} \int_{-\infty}^{\infty} dt \;
  e^{- i (\mathbf{q}\cdot \mathbf{R} - \omega t) }
  \langle 
     S^{\alpha}_{\mathbf{R}}(t)  S^{\alpha}_{\mathbf{0}} (0)
  \rangle \;,
\end{equation}
where $S^{\alpha}_{\mathbf{R}} (t)$ is the $\alpha=x,y,z$ component of spin operator at site $\mathbf{R}$ and time $t$. The $S^{\alpha\alpha}(\mathbf{q},\omega)$ is measured, among others, in inelastic neutron scattering, resonant inelastic X-ray scattering, electron spin resonance, and light absorption experiments. It gives precious information about the magnetic excitations in the material and the nature of the ground state, and helps to develop theoretical models.

% It is essential when we develop a theoretical model since it allows detailed comparison between the excitations of the model and the material, allowing us to discriminate between promising theories and dead ends.  

Unfortunately, the calculation of the dynamical properties in strongly correlated systems is notoriously difficult. Especially when frustration is present, analytical results are rare, and the available numerical methods are limited. 
For the calculation of ground state properties, a variational Monte Carlo (VMC) method based on Gutzwiller projected wave functions, pioneered for the SU(2) electron systems by Kaplan {\it et al.} \cite{Gutzwiller_Projection_1D_Heisenberg,*Horsch_1983} and enhanced in Refs.~\cite{1987JPSJ...56.1490Y,*1987JPSJ...56.3570Y,1987PhRvB..36..381G}, turned out to be very useful. 
Recently, based on a work by Li and Yang \cite{First_S_q_w_latter,*Yang_Li_2011PhRvB..83f4524Y}, Dalla Piazza {\it et al.} extended the variational Monte Carlo method to calculate the zero-temperature dynamical structure factor of the SU(2) Heisenberg model \cite{Excited_states_above_Fermi_sea}. 
The method constructs a finite-dimensional Hilbert space from the variational ground state, and Gutzwiller projected particle-hole excitations of the Fermi sea, and then evaluates the overlaps and the Hamiltonian matrix elements between these states by Monte Carlo sampling. 
The dynamical VMC is useful to examine the $S^{\alpha\alpha} (\mathbf{q},\omega)$  of spin liquids. 
It has been applied to the Heisenberg model on the kagome lattice \cite{Mei_Wen_arxiv_2015,2020PhRvB.102s5106Z}, and to the Heisenberg model with first- ($J_1$) and second-neighbor ($J_2$) interactions on one-dimensional chains \cite{Becca} and on square \cite{2018PhRvB..98m4410Y,2020PhRvB.102a4417F}, triangular \cite{2019PhRvX...9c1026F}, and honeycomb \cite{2020JPCM...32A4003F} lattices. 
Furthermore, it has been used to get spectral properties of correlated electrons in Refs.~\cite{PhysRevX.10.041023,2020PhRvB.101g5124I}.

The VMC also proved to be useful to characterize the ground state properties of the SU($N$) symmetric Mott insulators by introducing fermions with $N$-flavors (colors) \cite{2007JPCM...19l5215P}. 
The study of the SU(4) symmetric Heisenberg chain in the fundamental representation showed that the Gutzwiller projected Fermi sea of fermions with four flavors reproduced the critical exponents of the structure factor accurately \cite{2009PhRvB..80f4413W}. 
The method proved to be helpful to get insight into the properties of different two-dimensional SU(N) Heisenberg models showing spin-liquid behavior of different kinds \cite{SU4_Honeycomb,2016PhRvL.117p7202N,2020PhRvL.125k7202K}. 

Beyond pure theoretical interest, the SU($N$) symmetric Heisenberg models may realize in systems of ultracold atoms with fermionic statistics in optical lattices. Following initial theoretical proposals \cite{2003PhRvL..91r6402W,2009NJPh...11j3033C,2010NatPh...6..289G} and experiments \cite{2010PhRvL.105s0401T},
there were several reports about experimental observations of antiferromagnetic correlations in such systems \cite{Greif1307,2015Natur.519..211H,2016Sci...353.1257B,2018PhRvL.121v5303O,2020arXiv201007730T}.
Beside ultracold atoms, the spin-orbit coupled crystal field states in $\alpha-$ZrCl$_3$ may provide a material realization of an SU(4) Heisenberg model on the honeycomb lattice\cite{2018PhRvL.121i7201Y}, with an SU(4) spin liquid ground state having algebraic correlations \cite{SU4_Honeycomb}.

Given all this, it looks natural to adapt the dynamical VMC of Refs.~\cite{First_S_q_w_latter,*Yang_Li_2011PhRvB..83f4524Y,Excited_states_above_Fermi_sea,Mei_Wen_arxiv_2015,Becca} to SU($N$) symmetric Heisenberg models and calculate the dynamical structure factor. 
Here, we consider the one-dimensional SU(3) symmetric Heisenberg chain defined by the Hamiltonian
\begin{equation}
\mathcal{H} = J \sum_{i=0}^{L-1} \sum_{a=1}^{8} T_{i}^{a} T_{i+1}^{a} \;, \\
\label{eq:Hamilton}
\end{equation}
where $J$ is the exchange coupling, $L$ is the number of lattice sites, and $T_{i}^{a}$ are the 8 SU(3) spin operators acting on site $i$, with periodic boundary conditions $T_{0}^{a}\equiv T_{L}^{a}$. We extend the dynamical VMC to the SU(N) case and calculate the dynamical structure factor at zero temperature,
\begin{equation}
S^{aa}(k,\omega) = \sum_f | \langle f | T^{a}_{k} | 0 \rangle |^2 \delta(\omega-E_f+E_0) \;, \\
\end{equation}
where $|0\rangle$ is the ground state with energy $E_0$, the sum is over the $f$ excited states (each having energy $E_f$), and $k$ is the momentum.
We show that the particle-hole excitations of fermions with three colors describe the key features of the one-dimensional SU(3) symmetric Heisenberg model, including the central charge, the critical exponents, and the two-soliton continuum. 
Since the Gutzwiller projected Fermi sea is an exact eigenstate of the SU(3) Haldane-Shastry model \cite{1992PhRvB..46.1005K,1992PhRvB..46.9359H}, we used it to further benchmark our results. It turned out that some of the Gutzwiller projected particle-hole excited states are also exact eigenstates of the Haldane-Shastry model.

On the technical side, in the original papers of Li and Yang the importance sampling required a separate Monte Carlo simulation for each wave vector, in order to account for the weights of each particle-hole excitation \cite{First_S_q_w_latter,*Yang_Li_2011PhRvB..83f4524Y}. In later works Mei and Wen \cite{Mei_Wen_arxiv_2015}, and Ferrari {\it et al.} \cite{Becca} speeded up the sampling procedure by performing a single Monte Carlo simulation for each wave vector simultaneously, with the cost of worsening the statistics. Mei and Wen used the lowest energy variational state in the subspace of $S^z_{T} = 1$ as a guiding function, while Ferrari {\it et al.} used the approximating ground state, but both neglected the weights of the particle-hole excitations. We improved the method of Li and Yang, taking into account the weights of all particle-hole excitations in a single Monte Carlo simulation. Our method is slower than that of Mei and Wen, or Ferrari {\it et al.} since they are using a single state only, but for the same number of samples we get better statistics for the excited states.

The article is structured as follows. 
In Sec.~\ref{sec:SU3} we present the su(3) algebra and the SU(3) symmetric Heisenberg model.
We introduce the Gutzwiller projected Fermi sea $P_{\text{G}}|\text{FS}\rangle$ as a variational ground state of the SU(3) Heisenberg model in Sec.~\ref{sec:PGFS}, together with the SU(3) Haldane-Shastry model.
We calculate the structure factor  in Sec.~\ref{sec:static} and discuss the single mode approximation based on   $P_{\text{G}}|\text{FS}\rangle$ in Sec.~\ref{sec:SMA}, which we use to extract the velocity of excitations.  
We check the scaling of the ground state energy and give an estimate for the central charge in Sec.~\ref{sec:EGS}.
We devote Sec.~\ref{sec:Skw} to the dynamic structure factor: we describe the dynamical VMC method and apply it to the Heisenberg model and  the Haldane-Shastry model. We also compare the VMC calculation with the exact results both for the Heisenberg and for the Haldane-Shastry model. 
We conclude in Sec.~\ref{sec:Conlusion}. 
The paper ends with Appendices where we describe the SU(3) double-commutator (Appendix~\ref{sec:SUN_double_commutator}), the generalized eigenvalue problem (Appendix~\ref{sec:gen_eigenvalue}), the Monte Carlo importance sampling (Appendix~\ref{sec:Monte_Carlo_of_H_and_O}), and the method of error estimation (Appendix~\ref{sec:error estimation}).

%%%%%%%%%%%%%%%%%%%%%%%%%%%%%%%%%%%%%%%%%%%%%%%%%%%%%%%
\section{The SU(3) symmetric Heisenberg model}
\label{sec:SU3}
%%%%%%%%%%%%%%%%%%%%%%%%%%%%%%%%%%%%%%%%%%%%%%%%%%%%%%%

The su(3) algebra is defined by $8$ generators $T^{a}$, $a= 1,2,\dots,8$, satisfying the
\begin{equation}
  [T^a,T^b] = i f_{abc} T^c
  \label{eq:su3_com}
\end{equation}
commutations relation, where $f_{abc}$ are the structure constants of the algebra \cite{SUrels_SciPostPhysLectNotes.21}. 
The $T^a$ are $d\times d$ matrices when they act on the (local) Hilbert space spanned by the $d$ dimensional irreducible representation of the SU(3).  It is customary to refer to the irreducible representations of the SU(3) either by their Young tableaux or by their dimensions $d$  set in boldface, $\mathbf{d}$.
Conventionally, the $T^a$  are normalized such that 
\begin{subequations}
\begin{align}
\text{Tr}\, T^a &= 0 \\ 
\text{Tr}\, T^a T^b &= \frac{1}{2} \delta_{ab} 
\end{align}
\end{subequations}
The operators which commute with all the generators are called Casimir operators. The quadratic Casimir operator is
\begin{equation}
\label{Casimir operator}
 C_1 \equiv \mathbf{T} \cdot \mathbf{T} = \sum_{a=1}^8 T^a T^a.
\end{equation}
and there is an additional cubic Casimir operator $C_2$. 

The defining (also called fundamental) representation is three-dimensional ($d=3$) and is denoted by $\mathbf{3}$. It is identified with a Young-tableau, having a single box. The $T^a$ operators are represented by $3 \times 3$ traceless matrices
\begin{equation}
\label{generators of SU(3)}
 T^a = \frac{1}{2} \lambda_a,
\end{equation}
where $\lambda_a$ are the eight Gell-Mann matrices. 
The quadratic Casimir operator in the subspace of the defining representation $\mathbf{3}$ acts like
\begin{equation}
\label{C_1 in the fundamental irrep}
  C_1 = \frac{4}{3} \mathbf{1} \;.
\end{equation}

In this manuscript we consider Mott-insulating chains with singly occupied sites, where each site can host one fermionic particle of 3 possible colors $\mathsf{A}$, $\mathsf{B}$ and $\mathsf{C}$ . Thus, the one-particle states on each site belong to the defining (fundamental) representation $\mathbf{3}$. Using the Gell-Mann matrices we may construct site operators that act on the Hilbert space of these fermions as
\begin{equation}
  T^a_j = \frac{1}{2} \sum_{\mu,\nu} f^\dagger_{j,\mu} \lambda^a_{\mu,\nu} f^{\phantom{\dagger}}_{j,\nu} \;,
  \label{eq:TGellMann}
\end{equation}
where $f^\dagger_{j,\mu}$ creates and $f^{\phantom{\dagger}}_{j,\mu}$ annihilates a fermion with color $\mu \in \{ \mathsf{A}$, $\mathsf{B}, \mathsf{C}\}$ at site $j$. Since the $T^a_j$ operators conserve the fermions, they commute with the 
\begin{equation}
 n_j = \sum_{\mu} f^\dagger_{j,\mu} f^{\phantom{\dagger}}_{j,\mu} = f^{\dagger}_{j,\mathsf{A}} f^{\phantom{\dagger}}_{j,\mathsf{A}} + f^{\dagger}_{j,\mathsf{B}} f^{\phantom{\dagger}}_{j,\mathsf{B}} + f^{\dagger}_{j,\mathsf{C}} f^{\phantom{\dagger}}_{j,\mathsf{C}}
 \label{eq:nj}
\end{equation}
fermion number operator,
\begin{equation}
  [ T^a_j, n_{j'} ]  = 0 \;.
  \label{eq:comm_T_n}
\end{equation}
In particular, we will consider the correlation functions of the diagonal operator
\begin{equation}
 T^3_j = \frac{1}{2} \left(f^\dagger_{j,\mathsf{A}} f^{\phantom{\dagger}}_{j,\mathsf{A}} - f^\dagger_{j,\mathsf{B}} f^{\phantom{\dagger}}_{j,\mathsf{B}}  \right) 
\end{equation}
in the following. $T^3_j$ is equivalent to the $S^z_j$ operator for SU(2) acting on the S=1/2 spins when $\mathsf{A} \equiv |\uparrow\rangle$ and $\mathsf{B}\equiv|\downarrow\rangle$.

Let us also mention, that in the defining representation the permutation operator
\begin{equation}
  \mathcal{P}_{i,j} = \frac{1}{3} + 2 \mathbf{ T}_i \cdot \mathbf{ T}_j ,
\end{equation}
provides an alternative form to the Hamiltonian (\ref{eq:Hamilton}),
\begin{equation}
\mathcal{H} = \frac{J}{2} \sum_{i=0}^{L-1} \left( \mathcal{P}_{i,i+1} - \frac{1}{3} \right) ,
\label{eq:HamiltonPij}
\end{equation}
where  $\mathcal{P}_{i,i+1}$ exchanges the colors on sites $i$ and $i+1$,  $\mathcal{P}_{i,i+1} |\dots \alpha_i \beta_{i+1} \dots \rangle = |\dots \beta_i \alpha_{i+1} \dots \rangle $. Since the action of $\mathcal{P}_{i,i+1}$ is independent of the number of colors, for $N$ colors it defines the $SU(N)$ symmetric Heisenberg model. For $N=3$ it has been solved using Bethe Ansatz by Uimin \cite{1970JETPL..12..225U}, and for general $N$  by Lai \cite{1974JMP....15.1675L}  and in greater detail by Sutherland   \cite{1975PhRvB..12.3795S}. The SU(3) symmetric Heisenberg model is often referred to as Uimin-Lai-Sutherland model in the literature. The ground state is a massless phase, its low-energy critical properties are described by the SU(3)$_1$ Wess-Zumino-Witten model \cite{1984NuPhB.247...83K}. The correlations show a period tripling consistent with the gap closing at $k=0$ and $k=\pm 2\pi/3$, as also confirmed numerically \cite{1991PhRvB..4411836F}. Similarly to the SU(2) Heisenberg model \cite{1D_excitations}, the dynamical structure factor shows a continuum of soliton excitations, nicely revealed in recent numerical calculations \cite{DMRG_Binder_PRB2020}.

One can also consider models with higher dimensional local Hilbert space. For example, the model with the self-adjoint representation $\mathbf{8}$ shows Z$_3$ symmetry-protected topological phases  \cite{2014PhRvB..90w5111M}. Using SU(3) bosons instead of fermions one can construct a Haldane-gapped model with $\mathbf{10}$ \cite{2020PhRvL.125e7202G}. Valence bonds solids may also appear for models with higher dimensional irreducible representations \cite{2007PhRvB..75f0401G}.

%%%%%%%%%%%%%%%%%%%%%%%%%%%%%%%%%%%%%%%%%%%%%%%%%%%%%%%
\section{The Gutzwiller projected Fermi sea}
\label{sec:PGFS}
%%%%%%%%%%%%%%%%%%%%%%%%%%%%%%%%%%%%%%%%%%%%%%%%%%%%%%%

Kaplan {\it et al.} found that the Gutzwiller projected half-filled Fermi sea provides an excellent variational ground state for the SU(2) $S=1/2$ Heisenberg model \cite{Gutzwiller_Projection_1D_Heisenberg,*Horsch_1983}. They have shown that the nearest-neighbor correlation is only about  0.2\% off from the exact value and that the spin-spin correlation function decays inversely with the distance, reproducing the exact exponent. This latter has been confirmed  by the analytical evaluation of the correlations of the Gutzwiller projected wave functions in Ref.~\cite{1987PhRvL..59.1472G}. 

\subsection{The projected Fermi sea for SU(3)} 

%%%% BEGIN FIG %%%%%%%%%%%%%%%%%%%%%%%%%%%%%%%%%%%%%%%%%%%%%%%%%%%
\begin{figure}
\includegraphics[width=0.95\columnwidth]{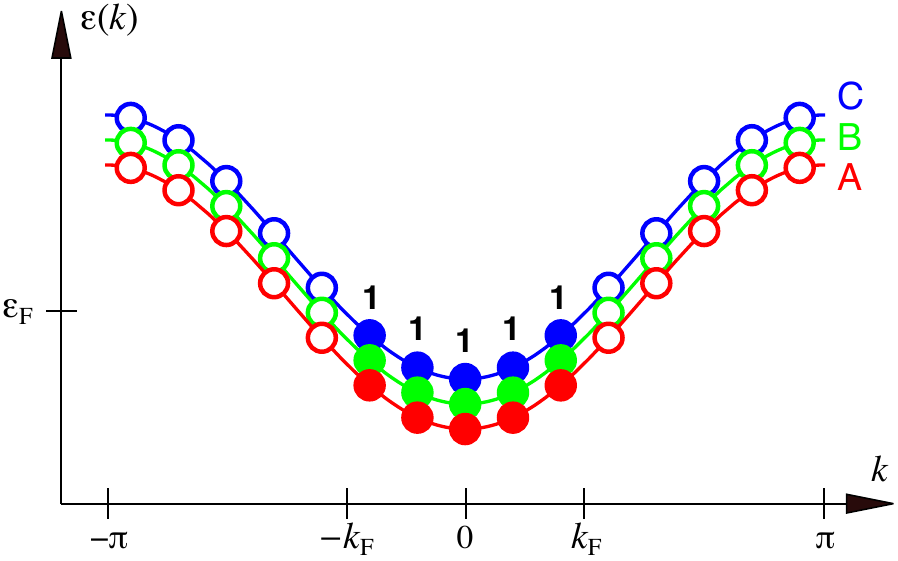}
\caption{
\label{fig:GS}
The filled Fermi sea of the fermions with 3 colors. 
The Fermi momentum is $k_F = \pi/3$, all the states between $-\pi/3$ and $\pi/3$ are occupied. At each $k$  the $f^\dagger_{k,\mathsf{A}} f^\dagger_{k,\mathsf{B}} f^\dagger_{k,\mathsf{C}} |0\rangle$ form an SU(3) singlet, denoted by $\mathbf{1}$. The Fermi seas of the three colors are degenerate, they were shifted for visualization.}
\end{figure}
%%%% END FIG %%%%%%%%%%%%%%%%%%%%%%%%%%%%%%%%%%%%%%%%%%%%%%%%%%%

This approach has been extended to the SU($N$) symmetric Heisenberg models in Refs.~\cite{2015PhRvB..91q4427D} and \cite{2009PhRvB..80f4413W}, showing that the Gutzwiller projected Fermi sea containing $N$ colors provides a good approximating ground state for the SU($N$) case as well. For $N=3$, the Gutzwiller projected Fermi sea is defined by
\begin{equation}
    P_{\text{G}} | \text{FS} \rangle = P_{\text{G}} \prod_{\alpha \in \{\mathsf{A},\mathsf{B},\mathsf{C}\}} \prod_{k \in \text{FS}} f^{\dagger}_{k,\alpha} | 0 \rangle,
    \label{eq:PGFS}
\end{equation}
where $| 0 \rangle$ is the vacuum, and the Gutzwiller projector is
\begin{equation}
    P_{\text{G}} = \prod_{i = 0}^{L - 1} \frac{n_i (n_{i} - 2) (n_{i} - 3)}{2} \;,
\end{equation}
$n_i$ being the fermion number operator defined in Eq.~(\ref{eq:nj}). Rewriting the Fermi sea to real space 
\begin{equation}
    | \text{FS} \rangle = \sum_{x} \text{det}( \{ R^{\mathsf{A}} \})  \text{det}(\{ R^{\mathsf{B}} \})  \text{det}(\{ R^{\mathsf{C}} \}) |x\rangle ,
    \label{eq:real_space_FS}
\end{equation}
where $|x \rangle = | \{ R^{\mathsf{A}} \} , \{ R^{\mathsf{B}} \} , \{ R^{\mathsf{C}} \} \rangle$,  $\{ R^{\mathsf{A}} \}$ are the lattice sites occupied by particles of color $\mathsf{A}$,  and $\text{det} ( \{ R^{\mathsf{A}} \} )$ is a Slater determinant of color $\mathsf{A}$ (and similarly for $\mathsf{B}$ and $\mathsf{C}$), which will be specified later. The Gutzwiller projector eliminates all configurations where any of the sites is not singly occupied. In the remaining configurations $|x\rangle$ each lattice site hosts one of the three fermionic particles $\mathsf{A}$, $\mathsf{B}$ or $\mathsf{C}$, and the number of particles of each color is equally $L/3$, with $L$ being the number of lattice sites. This is achieved at $1/3$ filling when the total number of fermions is equal to the number of sites, providing the $k_F=\pi/3$ Fermi momentum (Fig.~\ref{fig:GS}). The Slater determinant of color $\mathsf{A}$ is given by
\begin{equation}
    \text{det} ( \{ R^{\mathsf{A}} \} ) = 
    \begin{vmatrix}
    \xi_{1}(R^{\mathsf{A}}_{1}) & \xi_{1}(R^{\mathsf{A}}_{2}) &
    \ldots &
    \xi_{1}(R^{\mathsf{A}}_{L/3}) 
    \\
    \xi_{2}(R^{\mathsf{A}}_{1}) & \xi_{2}(R^{\mathsf{A}}_{2}) &
    \ldots &
    \xi_{2}(R^{\mathsf{A}}_{L/3}) 
    \\
    \vdots & \vdots & \ddots & \vdots 
    \\
    \xi_{L/3}(R^{\mathsf{A}}_{1}) & \xi_{L/3}(R^{\mathsf{A}}_{2}) &
    \ldots &
    \xi_{L/3    }(R^{\mathsf{A}}_{L/3})
    \end{vmatrix}
    ,
    \label{eq:slater_determinant}
\end{equation}
and similarly the Slater determinants of colors $\mathsf{B}$ and $\mathsf{C}$, where $\xi_{j}$ is the j-th lowest energy one-particle wavefunction of the non-interacting Hubbard Hamiltonian
\begin{equation}
    \mathcal{H} = -t \sum_{ i = 0}^{L - 1} f^\dagger_{i} f^{\phantom{\dagger}}_{i+1}.
    \label{eq:hopping_hamiltonian}
\end{equation}
Since this Hamiltonian is translationally invariant, these $\xi_{j}$ one-particle eigenstates can be chosen to be simultaneously eigenstates of the translation operator as well, with some eigenvalue $e^{iq}$, so that the eigenstates $\xi_{j}$ and the Slater determinants \ref{eq:slater_determinant} are complex. If the Fermi sea is non-degenerate, filling the lowest lying excited states results in filling pairs of wave vectors $q$ and $-q$ (Fig.~\ref{fig:GS}), which allows to make the states $\xi_{j}$ and the Slater determinants \ref{eq:slater_determinant} real by a suitable basis transformation. The boundary condition of the hopping Hamiltonian (\ref{eq:hopping_hamiltonian}) is chosen so as to make the Fermi sea non-degenerate, independently of the boundary condition of the original Heisenberg Hamiltonian (\ref{eq:Hamilton}), which is always periodic.

Since the $P_{\text{G}}$ is a function of the fermionic number operators, following Eq.~(\ref{eq:comm_T_n}) it commutes with the $T^a_j$,
\begin{equation}
  \left[ P_{\text{G}}, T^a_j \right] = 0,
\end{equation}
and so with the Casimir operator (\ref{Casimir operator}). Consequently, the projected wave function inherits the SU(3) quantum numbers of the unprojected, free fermion, wave function and is in the same irreducible representation. As the nondegenerate Fermi sea is a singlet, the Gutzwiller projected Fermi sea is also a singlet. To construct the singlet SU(3) Fermi sea, the number of particles (and so the number of sites) should be the multiple of 3.

\subsection{The SU(3) symmetric Haldane-Shastry model}

 Haldane \cite{1988PhRvL..60..635H} and Shastry \cite{1988PhRvL..60..639S} proved that the projected wave function is in fact the exact ground state of an SU(2) Heisenberg model with long range exchange interaction
\begin{equation}
  J_{i-j} = \frac{\pi^2 }{L^2 \sin^2 \frac{\pi (i-j)}{L}} \;,
  \label{eq:JlHS}
\end{equation}
proportional to the inverse squared chord distance between the spins at sites $i$ and $j$ arranged on a circle. Refs.~\cite{1992PhRvB..46.1005K,1992PhRvB..46.9359H} generalized the model to SU($N$) spins.
 The SU(3) symmetric Haldane-Shastry model is defined by 
\begin{equation}
\mathcal{H}_{\text{HS}} =  \sum_{i=0}^{L-2}\sum_{j=i+1}^{L-1} J_{i-j}\mathbf{T}_{i} \cdot \mathbf{T}_{j} \;,
\label{eq:HamiltonHSSU3}
\end{equation}
and it is the parent Hamiltonian of the projected SU(3) Fermi sea shown in Eq.~(\ref{eq:PGFS}). The dynamical properties of this model were studied in Refs.~\cite{2000PhRvL..84.1308Y,2000JPSJ...69..900Y,2005EL.....71..987S,2006PhRvB..73w5105S,2007PhRvL..98w7202G}.

%%%%%%%%%%%%%%%%%%%%%%%%%%%%%%%%%%%%%%%%%%%%%%%%%%%%%%%
\section{Structure factor}
\label{sec:static}
%%%%%%%%%%%%%%%%%%%%%%%%%%%%%%%%%%%%%%%%%%%%%%%%%%%%%%%

%%%%%%%%%%%%%%%%%%%%%%%%%%%%%%%%%%%%%%%%%%%%%%%%%%%%%%%
\begin{figure}[t]
	\centering
	    \includegraphics[width=0.9\columnwidth]{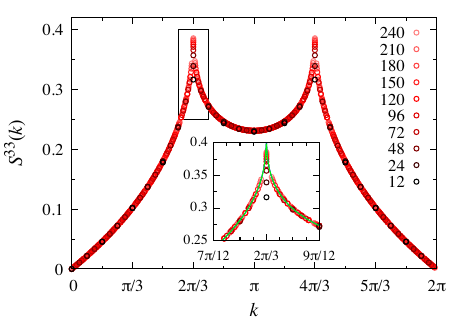}
	\caption{ 
	 The structure factor for chain lengths from $L=12$ to $L=240$. Except for the singular peaks at $k = \pm 2\pi/3$, all the points for different sizes fall onto a single curve.
The inset shows the power-law behavior of the $k = \pm 2\pi/3$ peaks.  Fitting a function of the form $a + b | k-2\pi/3|^{c} + d(k-2\pi/3)$ to the points around the peak (but excluding the peak itself)  gives an exponent $c = 0.337$ (green line). The value of the exponent is sensitive to the fitting window, it fluctuates around the correct $\eta -1 = 1/3$ value. The error bars are smaller than the symbol sizes. 
}
	\label{fig:Sk}
\end{figure}
%%%%%%%%%%%%%%%%%%%%%%%%%%%%%%%%%%%%%%%%%%%%%%%%%%%%%%%

The structure factor (spin-spin correlation function) of the SU($N$) Heisenberg model was calculated by quantum Monte Carlo technique in Refs.~\cite{1999PhRvL..82..835F,2012PhRvL.109t5306M} and by VMC in Refs.~\cite{2015PhRvB..91q4427D}. In this section we review the structure factor of the one-dimensional SU(3) symmetric Heisenberg model as calculated from the $P_{\text{G}} |\text{FS}\rangle$ and explore its critical properties. The  structure factor is the $\omega$-integrated dynamical structure factor
\begin{equation}
 S^{33}(k)  = \int_0^\infty d\omega \, S^{33}(k,\omega)  = \langle 0 | T^3_{-k} T^3_{k} | 0 \rangle    
 \label{eq:Sk}
\end{equation}
and depends only on the ground state $|0 \rangle$.  Here 
\begin{equation}
T^{a}_{k} = \frac{1}{\sqrt{L}} \sum_{j} e^{i k j} T^{a}_j
\end{equation} 
is the spin operator in the momentum representation.
In the following we will use the shorthand notation $\langle \ldots \rangle \equiv \langle 0 | \ldots | 0 \rangle$ for ground state averages. The structure factor is the Fourier transform of the static real-space correlation function
\begin{equation}
S^{33}(k) = \sum_j e^{i k j} \langle T_0^3 T_{j}^3 \rangle \;,
\label{eq:S33k_as_F_T_of_S33j}
\end{equation}
and it obeys the 
\begin{equation}
\label{eq:S33ksumrule}
\frac{1}{L} \sum_k S^{33}(k)  =  \langle T_0^3 T_0^3 \rangle = \frac{1}{8} \langle \mathbf{T}_0 \cdot \mathbf{T}_0 \rangle  =  \frac{1}{6} \;,
\end{equation}
 sum rule, 
where we used that $\mathbf{T}_0 \cdot \mathbf{T}_0$ equals the Casimir operator $C_1 = \frac{4}{3} \mathbf{1}$ in the fundamental representation, Eq.~(\ref{C_1 in the fundamental irrep}).

%%%%%%%%%%%%%%%%%%%%%%%%%%%%%%%%%%%%%%%%%%%%%%%%%%%%%%%
\begin{figure}[t]
	\centering
	    \includegraphics[width=0.9\columnwidth]{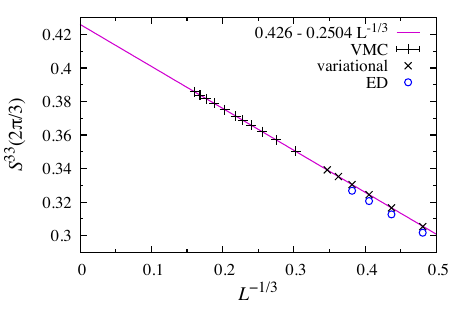}
	\caption{Finite size scaling of the $k=2\pi/3$ singular peak $S^{33}(2\pi/3)$, calculated from the Gutzwiller projected Fermi sea, plotted against $L^{-1/3}$. The straight line corroborates the non-analytic contribution proportional to $L^{-1/3}$.  }
	\label{fig:peaks_power}
\end{figure}
%%%%%%%%%%%%%%%%%%%%%%%%%%%%%%%%%%%%%%%%%%%%%%%%%%%%%%%

Fig.~\ref{fig:Sk} shows $S^{33}(k)$ obtained from the static real space correlation function as in Eq.~(\ref{eq:S33k_as_F_T_of_S33j}), which was calculated by using $P_{\text{G}} |\text{FS}\rangle$ as an approximating ground state 
\begin{equation}
   \langle T_0^3 T_{j}^3 \rangle  \approx \frac{ \langle \text{FS} | P_{\text{G} } T^{3}_0 T^{3}_j P_{\text{G}} | \text{FS} \rangle }{\langle \text{FS} | P_{\text{G} } P_{\text{G}} | \text{FS} \rangle }\;.
\end{equation}
We evaluated the equation above for small system sizes $L \leq 24$ exactly, and for $L > 24$ with Monte Carlo sampling of the approximating ground state. The error bars for most of the measured quantities related to static correlations were smaller than the symbol sizes. Details of the error estimation can be found in Appendix \ref{sec:error estimation}.

The critical theory of the SU(3) Heisenberg model is the SU(3)$_1$ Wess-Zumino-Witten model \cite{1984NuPhB.247...83K}, and the singularity at $k=\pm 2\pi/3$ in $S^{33}(k)$  can be traced back to the oscillating algebraic decay of the correlation function 
\begin{equation}
\langle \mathbf{T}_{j} \cdot \mathbf{T}_{j'} \rangle \propto \frac{1}{(j-j')^2} + 
\frac{1}{|j-j'|^\eta} \cos \left[\frac{2 \pi}{3} (j- j')\right]
\label{eq:TiTj}
\end{equation}
where the exponent is \cite{1986NuPhB.265..409A}
\begin{equation}
  \eta =  \frac{4}{3} \;.
\end{equation}
More detailed renormalization group analysis revealed  logarithmic corrections in the correlation function \cite{1997PhRvB..55.8295I}. The critical properties were confirmed by DMRG method in  Refs. \cite{2008AnP...520..922F,2009PhRvB..79a2408A} and QMC in Ref.~\cite{2012PhRvL.109t5306M}.

As a consequence of the algebraic decay, Eq.~(\ref{eq:TiTj}), the Fourier transform of the correlation function will show a power-law singularity at $k=\pm 2\pi/3$, 
\begin{align}
\label{power law behaviour of the static peak}
S^{33}(\pm 2\pi/3 + \delta k) &\propto \left| \delta k \right|^{\eta-1} \propto \left| \delta k \right|^{\frac{1}{3}} \;.
\end{align}
The singularity at the $k=\pm 2 \pi/3$ and the power law like behaviour in its vicinity is clearly seen in the inset of Fig.~\ref{fig:Sk}. To extract more precisely the behavior of the singular peaks at $k=\pm2\pi/3$, we follow \cite{1990PhRvB..41.2326O}: the exponent controls the non-analytical finite size behavior, as it should go with $\propto L^{-\frac{1}{3}}$. Fig.~\ref{fig:peaks_power} confirms our expectations, the 
$S^{33}( 2\pi/3)$ clearly has a component that is linear in $L^{-\frac{1}{3}}$.

%%%%%%%%%%%%%%%%%%%%%%%%%%%%%%%%%%%%%%%%%%%%%%%%%%%%%%%
\section{Single mode approximation and the velocity of excitations}
\label{sec:SMA}
%%%%%%%%%%%%%%%%%%%%%%%%%%%%%%%%%%%%%%%%%%%%%%%%%%%%%%%

%%%%%%%%%%%%%%%%%%%%%%%%%%%%%%%%%%%%%%%%%%%%%%%%%%%%%%%
\begin{figure}[b]
	\centering
\includegraphics[width=0.9\columnwidth]{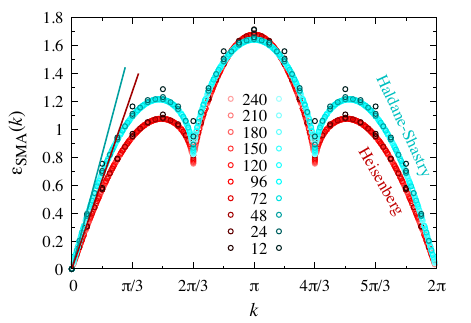}
	\caption{ 
	The energy of excitations in the single mode approximation, $\varepsilon_{\text{SMA}}(k)$, for the SU(3) Heisenberg model (red points) and the Haldane-Shastry model (blue points) calculated from the Gutzwiller projected Fermi sea for system sizes ranging from 12 to 240. The lines show the velocities of excitations, obtained from fitting the SMA in the $k\to 0$ limit. The error bars (not shown) are smaller than the symbol sizes.}
		\label{fig:SMA}
\end{figure}
%%%%%%%%%%%%%%%%%%%%%%%%%%%%%%%%%%%%%%%%%%%%%%%%%%%%%%%

%%%%%%%%%%%%%%%%%%%%%%%%%%%%%%%%%%%%%%%%%%%%%%%%%%%%%%%
\begin{figure}[t]
	\centering
	    \includegraphics[width=0.9\columnwidth]{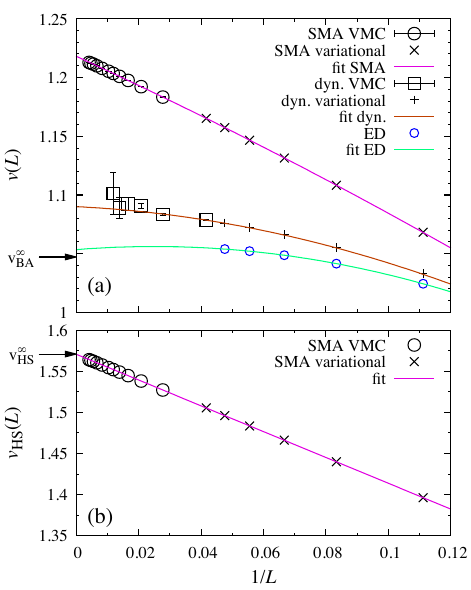}

	\caption{ 
	Finite size scaling of the velocities $v(L) = \varepsilon(k_{\text{min}})/k_{\text{min}}$. (a) The velocities of the SU(3) symmetric Heisenberg model obtained from exact diagonalization (ED), single mode approximation (SMA, calculated from the Gutzwiller projected Fermi sea used as an approximating ground state), and the variational dynamical structure factor calculated by using particle-hole excitations. The arrow  represents the exact Bethe Ansatz result in the thermodynamic limit, $v_{\text{BA}}^{\infty} = \pi/3$.
	 (b) The single mode approximation using the Gutzwiller projected Fermi sea gives the exact value of the velocity for the Haldane-Shastry model, $v_{\text{HS}}^{\infty} = \pi/2$ for $L\to \infty$ (denoted by the arrow).}
	\label{fig:v_finite_size}
\end{figure}
%%%%%%%%%%%%%%%%%%%%%%%%%%%%%%%%%%%%%%%%%%%%%%%%%%%%%%%

The single mode approximation (SMA) assumes that the dynamical structure factor consists of a single excitation created by acting some momentum-dependent operator (e.g. density) on the ground state \cite{1953PhRv...91.1291F,1986PhRvB..33.2481G}. 
Actually, since the dynamical structure factor  of the SU(3) Heisenberg model consist of two- and multi-soliton continua \cite{1975PhRvB..12.3795S,2000PhRvL..84.1308Y,2005EL.....71..987S,DMRG_Binder_PRB2020}, we shall not expect the SMA to work in general. However, the two-soliton continuum narrows at small momenta, and the SMA allows to extract the velocity of the excitations. Furthermore, it helps to check the reliability of the variational approach when we calculate the dynamical structure factor in Sec.~\ref{sec:Skw} below.
%\noteKP{compare SMA with the first moment}
%

\subsection{The Heisenberg model}

We create the excitation by applying $T^3_k$ to the $  P_{\text{G}} | \text{FS} \rangle$. 
The energy of this excitation is
\begin{equation}
 \varepsilon_{\text{SMA}}(k) = \frac{f(k)}{S^{33}(k)}
\end{equation}
where $f(k)$ is the oscillator strength defined as the first moment of the dynamical structure factor, and $S^{33}(k)$ is the structure factor defined in Eq.~(\ref{eq:Sk}). 
The oscillator strength $f(k)$ can be expressed using a double commutator, and we get
\begin{equation}
  \varepsilon_{\text{SMA}}(k) = \frac{1}{2} \frac{\langle 0 | \left[[T^3_{-k},\mathcal{H}],  T^3_{k} \right] | 0 \rangle}{\langle 0 | T^3_{-k} T^3_{k} | 0 \rangle} \;.
\end{equation}
We calculate the double commutator in Appendix \ref{sec:SUN_double_commutator}. For the one-dimensional SU(3) symmetric Heisenberg model, following Eq.~(\ref{eq:f(q)1D}), the oscillator strength becomes
\begin{equation}
    f(k) = -
     6 J \sin^2 \frac{k}{2} \left\langle T_0^3 T_1^3 \right\rangle \;.
\end{equation}
The energy of the excitation in the SMA is then given by
\begin{equation}
 \varepsilon_{\text{SMA}}(k) = 
 - 6 J \sin^2 \frac{k}{2}  \frac{  \left\langle T_0^3 T_1^3 \right\rangle}{\sum_j e^{i k j} \langle T_0^3 T_{j}^3 \rangle}
\end{equation}
The red circles in Fig.~\ref{fig:SMA} show $\varepsilon_{\text{SMA}}(k)$ calculated for system sizes up to $L=240$, using the Gutzwiller projected Fermi sea as an approximating ground state. The velocity of an excitation is the slope of the energy of the excitation in the  $k\to 0$ limit,  i.e. $\varepsilon_{\text{SMA}}(\delta k) \approx v \delta k$, assuming that the mode is well defined, which turns out to be the case in the long-wavelength limit. We may therefore use the SMA, calculated from the Gutzwiller projected Fermi sea, to extract the velocity of low-energy excitations, as shown in Fig.~\ref{fig:v_finite_size}(a) for the Heisenberg model.
% Up to $L = 24$ the velocities were calculated exactly (meaning without statistical error, since we are using an approximating ground state), while for larger system sizes up to $L = 240$ we used Monte Carlo evaluation. In the same figure we also show the velocities obtained from exact diagonalization (up to $L = 18)$, and the dynamical structure factor presented in the next section (up to $L = 72$, again no statistical error until $L = 21)$. 
The fit 
\begin{equation}
\label{eq:fit_ED_for_velocity}
 v_{\text{ED}}(L) = v_{\text{ED}}^{\infty}  + b_{\text{ED}} L^{-1} + c_{\text{ED}} L^{-2} \;,
\end{equation}
 for velocities obtained from ED gives $v_{\text{ED}}^{\infty} = 1.0535 \pm 2 \cdot 10^{-4}$, $b_{\text{ED}} = 0.213 \pm 0.005$ and $c_{\text{ED}} = -4.27 \pm 0.03$, where the errors come from the covariance matrix of the fit. The exact result for the velocity 
\begin{equation}
  v_{\text{BA}} = \frac{\pi}{3} \approx 1.0472 
  \label{eq:vBA}
\end{equation}
is known from the Bethe Ansatz \cite{1975PhRvB..12.3795S}. The relative error of the $v_{\text{ED}}^{\infty}$ is 0.6\%, which is about 30$\times$ larger than the error estimated from the covariance matrix. This suggests that the fitting form Eq.~(\ref{eq:fit_ED_for_velocity}) is unlikely the true form of the finite size scaling (for example, the ground state energy has corrections logarithmic in system size \cite{1997PhRvB..55.8295I}).
 We note that our ED estimate for the velocity agrees with the $v = 1.0535$ obtained by DMRG in Ref.~\cite{2015PhRvB..92g5128S} (see also \cite{2015PhRvL.114n5301C} where $v=1.2643$ for a more complicated $S=2$ spin model with emerging SU(3) symmetry).

We fitted a quadratic polynomial on the velocities obtained from the SMA, of the form
\begin{equation}
\label{velocity from SMA of the Gutzwiller projected Fermi sea}
  v_{\text{SMA}}(L) = v_{\text{SMA}}^{\infty}  + b_{\text{SMA}} L^{-1} + c_{\text{SMA}} L^{-2} \;,
\end{equation}
with $v_{\text{SMA}}^{\infty} \approx 1.21822 \pm 3 \cdot 10^{-5}$, $b_{\text{SMA}} \approx -1.227 \pm 0.002$, and $c_{\text{SMA}} \approx -1.01 \pm 0.01$. The extrapolated velocity  $v_{\text{SMA}}^\infty$ is therefore about $16\%$ larger than $v_{\text{BA}}$, the exact value.

A better approximation can be obtained, if the velocity is extracted from the dynamical structure factor calculated approximately using particle-hole excitations, as explained later in Sec.~\ref{sec:Method}. 
Fitting a quadratic polynomial
\begin{equation}
\label{velocity from dyn VMC}
  v_{\text{dyn}}(L) = v_{\text{dyn}}^{\infty} + b_{\text{dyn}} L^{-1}  +  c_{\text{dyn}} L^{-2}  \;,
\end{equation}
yields $v_{\text{dyn}}^{\infty} \approx 1.0901 \pm 8 \cdot 10^{-4}$,  $b_{\text{dyn}} \approx -0.13 \pm 0.02$, and $c_{\text{dyn}} \approx -3.5 \pm 0.15$. In this approximation the velocity is much closer to the exact value, with an error about $4 \%$.

\subsection{Haldane-Shastry model}
It is quite instructive to apply the SMA to the Haldane-Shastry model. The Gutzwiller projected Fermi sea is the exact ground state wave function of the $\mathcal{H}_{\text{HS}}$ (\ref{eq:HamiltonHSSU3}), therefore the $\varepsilon_{\text{SMA}}(k)$ provides a variational upper bound on the energy of the excitations. Inserting the long-range $J_l$ of the Haldane-Shastry model, Eq.~(\ref{eq:JlHS}), into 
Eq.~(\ref{eq:f(q)1D}), we get
\begin{equation}
    f^{\text{HS}}(k) = -3 \frac{\pi^2 }{L^2} \sum_{l=1 }^{L-1} \frac{\sin^2 \frac{k l }{2} }{\sin^2 \frac{\pi l }{L}} \left\langle T^3_0 T^3_l \right\rangle \;.
    \label{eq:fk_HS}
\end{equation}
The $\varepsilon_{\text{SMA}}(k)$ calculated numerically from the expression above and the $S^{33}(k)$ is plotted in Fig.~\ref{fig:SMA} with blue circles. It resembles very much to that of the Heisenberg model, they are both gapless at $k\to 0$ and show a finite gap at the critical $k=2\pi/3$, where we expect a gapless continuum. Not surprisingly, the SMA is unable to capture the vanishing gap of the 2-coloron continuum. 

The oscillator strength becomes trivial for the smallest value of the momentum, $k_{\text{min}}=2\pi/L$, as the sines cancel in Eq.~(\ref{eq:fk_HS}):
\begin{equation}
    f^{\text{HS}}(k_{\text{min}}) = 
    -3 \frac{\pi^2 }{L^2} \sum_{l=1 }^{L-1}  \left\langle T^3_0 T^3_l \right\rangle
    = \frac{\pi^2 }{2 L^2 } \;,
\end{equation}
where we used that $\sum_{l=1}^{L-1} \left\langle T^3_0 T^3_l\right\rangle = -\left\langle T^3_0 T^3_0 \right\rangle = -1/6$, since $\sum_{l=0}^{L-1} \left\langle T^3_0 T^3_l\right\rangle = 0$ in the singlet ground state.
The exact value of the correlation function for the smallest momentum,
\begin{equation}
S^{33}(k_{\text{min}}) = \frac{1}{2L-2} \;,
\end{equation}
 is known from \cite{2000JPSJ...69..900Y}, and is also obeyed by our $S^{33}(k)$ data. Therefore the exact SMA energy at $k_{\text{min}}$ is
 \begin{equation}
\varepsilon^{\text{HS}}_{\text{SMA}}(k_{\text{min}}) = \frac{f^{\text{HS}}(k_{\text{min}}) }{S^{33}(k_{\text{min}})} = \pi^2 \frac{L-1}{ L^2 } \;,
\end{equation}
and for the velocity we get
\begin{equation}
  v_{\text{HS}}(L) 
  = \frac{\varepsilon_{\text{SMA}}^{\text{HS}}(k_{\text{min}})}{k_{\text{min}}} = \frac{\pi}{2} \left(1-\frac{1}{L} \right) .
  \label{eq:vHSL} 
\end{equation}
The SMA recovers the exact  $v_{\text{HS}} = \pi/2$ in the thermodynamic limit \cite{2000PhRvL..84.1308Y}. The $v_{\text{HS}}(L)$ is shown as a straight line through the points obtained by numerically exact calculation of the SMA (for $L \leq 24)$ and SMA evaluated by Monte Carlo (for $24 < L \leq 240$) in Fig.~\ref{fig:v_finite_size}(b). 

%%%%%%%%%%%%%%%%%%%%%%%%%%%%%%%%%%%%%%%%%%%%%%%%%%%%%%%
\section{Scaling of the ground state energy}
\label{sec:EGS}
%%%%%%%%%%%%%%%%%%%%%%%%%%%%%%%%%%%%%%%%%%%%%%%%%%%%%%%

%%%%%%%%%%%%%%%%%%%%%%%%%%%%%%%%%%%%%%%%%%%%%%%%%%%%%%%
\begin{figure}[t]
	\centering
	    \includegraphics[width=0.9\columnwidth]{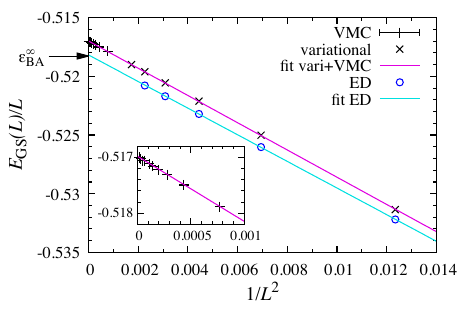}
	\caption{ 
	Finite size scaling of the ground state energy density, calculated for the approximating variational ground state $P_{\text{G}}|\text{FS} \rangle$ (variational and VMC denotes the points calculated numerically exactly and with Monte Carlo, respectively), and compared with exact diagonalization results (ED). The arrow shows the exact energy density in the thermodynamic limit obtained from Bethe Ansatz, Eq.~(\ref{eq:eBA}).}	\label{fig:EGS_finite_size}.
\end{figure}
%%%%%%%%%%%%%%%%%%%%%%%%%%%%%%%%%%%%%%%%%%%%%%%%%%%%%%%

According to the conformal theory \cite{1984JPhA...17L.385C,1986PhRvL..56..742B,1986PhRvL..56..746A}
, the finite-size scaling of the ground state energy,
\begin{align}
  E(L) = L \varepsilon^{\infty} - \frac{\pi}{6L} v c \;,
  \label{eq:E(L)}
\end{align}
supplies information about the central charge $c$ and the velocity $v$ of the excitations.
The SU(3)$_1$ Wess-Zumino-Witten model, the critical theory of the Heisenberg model, has a central charge $c = 2$. The ground state  energy density from the Bethe Ansatz solution is 
\begin{equation}
  \varepsilon^{\infty} = \varepsilon^{\infty}_{\text{BA}} = \frac{1}{3} -\frac{\pi }{6 \sqrt{3}} - \frac{\ln 3}{2} \approx -0.518273 \;
  \label{eq:eBA} 
\end{equation}
in the thermodynamic limit\cite{1975PhRvB..12.3795S}, and the velocity is given by Eq.~(\ref{eq:vBA}). Let us now check to what extent is Eq.~(\ref{eq:E(L)}) reproduced by the projected wave function.

To this end, we plotted the ground state energy density $E(L)/L$ vs. $1/L^2$ for several system sizes in Fig.~\ref{fig:EGS_finite_size}. 
Fitting a function $ b - a /L^2 $ to the ground state energy densities $E(L)/L$ obtained from the Gutzwiller projected Fermi sea has given $\varepsilon^{\infty} = b \approx -0.516981 \pm 2 \cdot 10^{-6}$ and $\frac{\pi}{6} vc = a \approx 1.1612 \pm 3 \cdot 10^{-4}$, so $ vc \approx 2.2178 \pm 6 \cdot 10^{-4}$, which compared to the exact  $v_{\text{BA}} c =  2\pi/3  \approx 2.0944$ is within $6\%$. It shows that the Gutzwiller projected Fermi sea gives a good approximation for the product of the velocity and the central charge. However, the velocity (\ref{velocity from SMA of the Gutzwiller projected Fermi sea}) calculated from the SMA of the Gutzwiller projected Fermi sea was $v_{\text{SMA}}^\infty \approx 1.21822 \pm 3 \cdot 10^{-5}$ instead of $v_{\text{BA}} \approx 1.0472$, therefore the central charge calculated solely from the Gutzwiller projected Fermi sea is less precise, $c \approx 1.7192 \pm 5 \cdot 10^{-4}$.

A better estimate of the central charge can be achieved using the velocity $v_{\text{dyn}}^{\infty} \approx 1.0901 \pm 8 \cdot 10^{-4}$ extracted from the fit (\ref{velocity from dyn VMC}). This gives $c \approx 2.034 \pm 0.002$, which is within an error of $2 \%$. For the details of error estimation see Appendix \ref{sec:error estimation}.

In comparison, the fit to the ground state energy we got from exact diagonalization of $L=9, 12,15,18,$ and $21$ gives  $E(L)/L= -0.518186 - 1.13295/L^2$, so  $v c = 2.16377$ is closer to the   $v_{\text{BA}} c$, but not yet there. The reason for the poor agreement is due to the logarithmic corrections \cite{1997PhRvB..55.8295I}.

%%%%%%%%%%%%%%%%%%%%%%%%%%%%%%%%%%%%%%%%%%%%%%%%%%%%%%%
\section{Dynamical structure factor}
\label{sec:Skw}
%%%%%%%%%%%%%%%%%%%%%%%%%%%%%%%%%%%%%%%%%%%%%%%%%%%%%%%

In this section we calculate the dynamical structure factor at zero temperature, defined by
\begin{equation}
\label{dynamical structure factor}
     S^{ab}(\mathbf{k},\omega) = 2\pi \sum_{\lambda} \langle 0 | T^{a}_{\mathbf{-k}}| \lambda \rangle \langle \lambda | T^{b}_{\mathbf{k}}| 0 \rangle  \delta(\omega + E_{0} - E_{\lambda}),
\end{equation}
where $| \lambda \rangle $ are eigenstates of $\mathcal{H}$ with energies $E_{\lambda}$, and $T^{a}_{\mathbf{k}} = \frac{1}{\sqrt{L}} \sum_{\mathbf{R}} e^{i \mathbf{k} \cdot \mathbf{R}} T^{a}_{\mathbf{R}}$. Because of the SU(3) rotational symmetry of the Heisenberg Hamiltonian in spin space, off-diagonal terms with $a \neq b$ vanish, and all eight diagonal components are equal,
\begin{equation}
\label{diagonal structure factors}
    S^{11}(\mathbf{k},\omega) = S^{22}(\mathbf{k},\omega) = \ldots = S^{88}(\mathbf{k},\omega).
\end{equation}
Since the $T^3$ and $T^8$ are diagonal, calculating $S^{33}(\mathbf{k},\omega)$ and $S^{88}(\mathbf{k},\omega)$ requires eigenstates $| \lambda \rangle$ which have $T^3_{\text{total}} = 0$ and $T^8_{\text{total}} = 0$ just like the ground state (similarly to $S^z_{\text{total}} = 0$ in case of SU(2)), thus calculating $S^{33}(\mathbf{k},\omega)$ and $S^{88}(\mathbf{k},\omega)$ is easier than that of the non-diagonal $T^a_{\mathbf{k}}$ operators. 

\subsection{Method}
\label{sec:Method}

%%%% BEGIN FIG %%%%%%%%%%%%%%%%%%%%%%%%%%%%%%%%%%%%%%%%%%%%%%%%%%%
\begin{figure}[bt!]
\centering
\includegraphics[width=0.95\columnwidth]{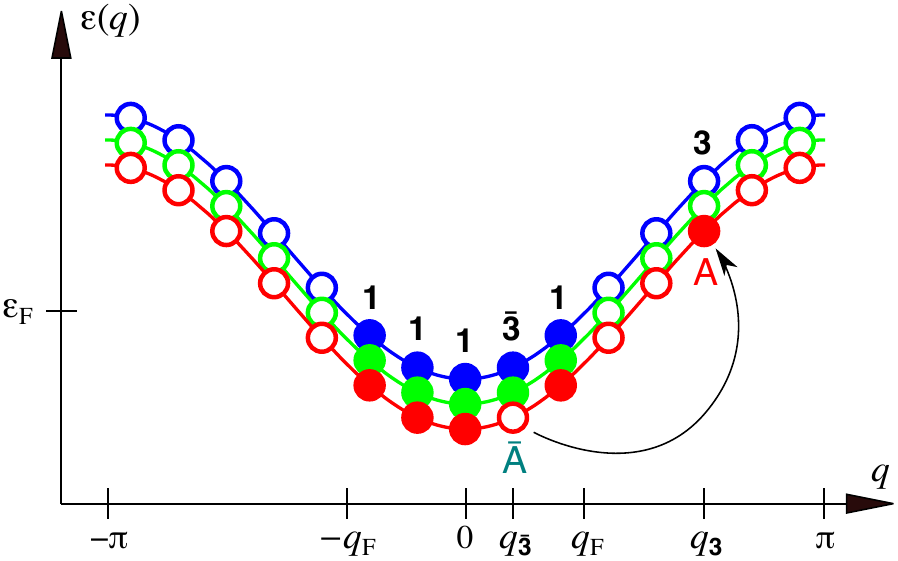}
\caption{
 The particle-hole excitation $|k,q_{\mathbf{\bar 3}},\mathsf{A} \rangle = P_{\text{G}} f^{{\dagger}}_{q_{\mathbf{ 3}},\mathsf{A}} f^{\phantom{\dagger}}_{q_{\mathbf{\bar 3}},\mathsf{A}} |\text{FS} \rangle$ 
 with $k=q_{\mathbf{ 3}} - q_{\mathbf{\bar 3}}$,  $q_{\mathbf{\bar 3}} \in \text{FS}$ and  $q_{\mathbf{ 3}}  \not\in \text{FS}$.
 Such particle-hole excitations span the truncated Hilbert space.
  The states 
  $|k,q_{\mathbf{\bar 3}},\mathsf{A} \rangle$, 
  $|k,q_{\mathbf{\bar 3}},\mathsf{B} \rangle$, 
  and $|k,q_{\mathbf{\bar 3}},\mathsf{C} \rangle$ 
  correspond to the three states in the center of the weight diagrams of 
  $\mathbf{3} \otimes \overline{\mathbf{3}} = \mathbf{1} \oplus \mathbf{8}$ 
  which has all three particles of different colors. 
These are therefore linear combinations of the singlet state $\mathbf{1}$ of the form 
$\frac{1}{\sqrt{3}}\left(|k,q_{\mathbf{\bar 3}},\mathsf{A} \rangle+|k,q_{\mathbf{\bar 3}},\mathsf{B} \rangle+|k,q_{\mathbf{\bar 3}},\mathsf{C} \rangle \right)$ and the two states 
$\frac{1}{\sqrt{2}}\left(|k,q_{\mathbf{\bar 3}},\mathsf{A} \rangle - |k,q_{\mathbf{\bar 3}},\mathsf{B} \rangle \right)$ and 
$\frac{1}{\sqrt{6}}\left(|k,q_{\mathbf{\bar 3}},\mathsf{A} \rangle+|k,q_{\mathbf{\bar 3}},\mathsf{B} \rangle - 2|k,q_{\mathbf{\bar 3}},\mathsf{C} \rangle \right)$ 
belonging to the adjoint representation $\mathbf{8}$. 
The latter ones are used for the calculation of $S^{33}(k,\omega)$ and $S^{88}(k,\omega)$. The three dispersions are degenerate, but shifted for visualization.}
\label{fig:particle-hole excitation}
\end{figure}
%%%% END FIG %%%%%%%%%%%%%%%%%%%%%%%%%%%%%%%%%%%%%%%%%%%%%%%%%%%

%%%% BEGIN FIG %%%%%%%%%%%%%%%%%%%%%%%%%%%%%%%%%%%%%%%%%%%%%%%%%%%
\begin{figure*}[b!t]
\centering
\includegraphics[width=0.95\textwidth]{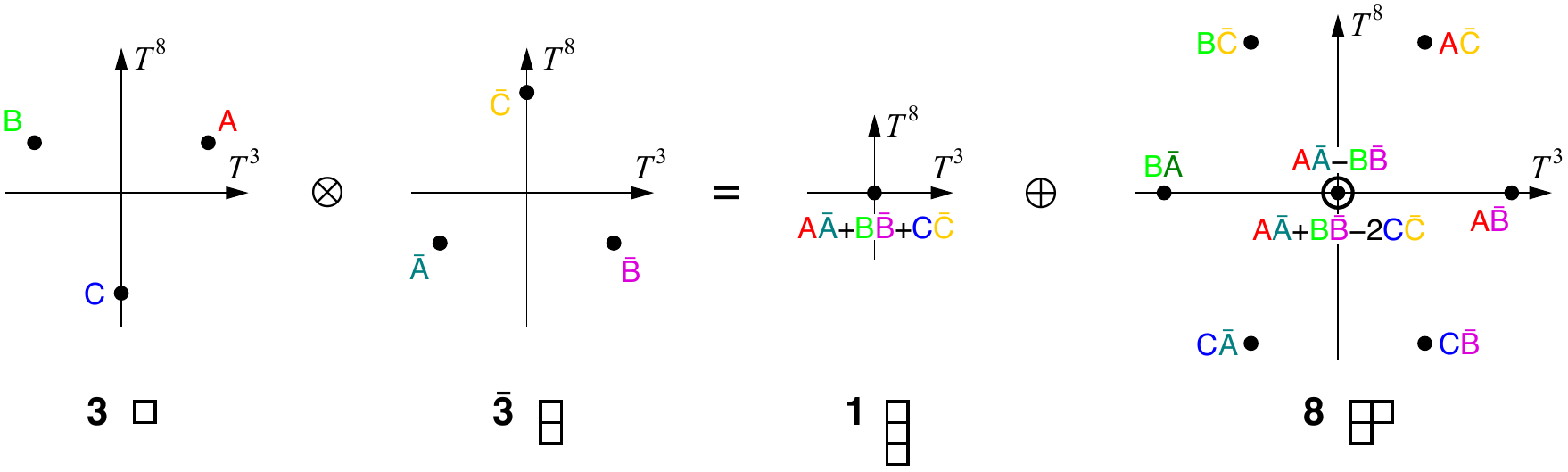}
\caption{Graphical representation of the $\mathbf{3} \otimes \mathbf{\bar 3} = \mathbf{1} \oplus \mathbf{8}$ decomposition of the one particle-hole excitations using the weight diagrams. The irreducible representation of a single particle is the three-dimensional $\mathbf{3}$, represented by a Young tableau with a single box. The weight diagram shows the three states, the values of the diagonal $T^3$ and $T^8$ operators for an added $\mathsf{A}$ fermion are $(T^3,T^8)=(1/2,1/2\sqrt{3})$ -- these are the coordinates of the point labeled by the red $\mathsf{A}$. The coordinates of the added $\mathsf{B}$ fermion are $(-1/2,1/2\sqrt{3})$, and of the $\mathsf{C}$ are $(0,1/\sqrt{3})$. The irreducible representation of a hole is $\mathbf{\bar 3}$ since the two remaining  fermions anti-symmetrize, this is denoted by two vertical boxes in the Young-tableau. 
The hole $\mathsf{\bar A}$ (colored by teal), with coordinates $(-1/2,-1/2\sqrt{3})$, is the anti-symmetrical combination of $\mathsf{B}$ and $\mathsf{C}$ fermions (see also Fig.~\ref{fig:particle-hole excitation}). The product of $\mathbf{3} \otimes \mathbf{\bar 3}$ contains a singlet, Eq.~(\ref{eq:kq1}), and the eight states of the $\mathbf{8}$. The two states of Eqs.~(\ref{eq:kR83}) and (\ref{eq:kR88}) are at the center $(T^3,T^8)=(0,0)$ of the weight diagram for $\mathbf{8}$. These constitute the final states in the $S^{33}(\mathbf{k},\omega)$ and $S^{88}(\mathbf{k},\omega)$ structure factor when the initial state is a singlet $\mathbf{1}$. }
\label{fig:weight_diagrams}
\end{figure*}
%%%% END FIG %%%%%%%%%%%%%%%%%%%%%%%%%%%%%%%%%%%%%%%%%%%%%%%%%%%

Following \cite{First_S_q_w_latter,*Yang_Li_2011PhRvB..83f4524Y,Excited_states_above_Fermi_sea,Becca} we calculate $S^{33}(\mathbf{k},\omega)$ by approximating the ground state with $P_{\text{G}} | \text{FS} \rangle$, and the excited states $| \lambda \rangle$ with approximating excited states $| \phi_n \rangle$. The $| \phi_n \rangle$ are found by building a set of particle-hole excited states upon the approximating ground state (which are not true eigenstates of $\mathcal{H}$), projecting the Hamiltonian onto these particle-hole excited states, and solving the generalized eigenvalue problem for this projected Hamiltonian $\tilde{\mathcal{H}}$. The $| \phi_n \rangle$ are then the eigenstates obtained from the generalized eigenvalue problem.
Since the subspace of particle-hole excited states is not closed under the action of the Hamiltonian $\mathcal{H}$, the eigenstates $| \phi_n \rangle$ of $\tilde{\mathcal{H}}$ are only approximating eigenstates of $\mathcal{H}$.
The main advantage of this method is that the number of particle-hole excited states grows as $\propto L^2$, while the dimension of the Hilbert-space grows exponentially in $L$. Solving the generalized eigenvalue problem provides the excitation energies directly.

%\subsubsection{Particle-hole excitations}

Building on the work of Dalla Piazza {\it et al.} \cite{Excited_states_above_Fermi_sea}, we construct the particle-hole excited states as 
\begin{equation}
\label{particle-hole excited states}
|\mathbf{k},\mathbf{q},\sigma \rangle = P_{\text{G}} f^{{\dagger}}_{\mathbf{k+q},\sigma} f^{\phantom{\dagger}}_{\mathbf{q},\sigma} |\text{FS} \rangle ,
\end{equation}
where $\sigma\in\{\mathsf{A},\mathsf{B},\mathsf{C}\}$ is the color of the fermion being moved from the Fermi sea ($ \mathbf{q} \in \text{FS}$) into an unoccupied state ($\mathbf{k+q} \not\in \text{FS}$), as illustrated in Fig.~\ref{fig:particle-hole excitation}. The excitation above does not change the number of fermions of different colors. For a fixed $\mathbf{k}$ and $\mathbf{q}$, the linear combination
\begin{equation}
|\mathbf{k},\mathbf{q},\mathbf{1} \rangle =
\frac{1}{\sqrt{3}}
\left(
|\mathbf{k},\mathbf{q},\mathsf{A} \rangle
+|\mathbf{k},\mathbf{q},\mathsf{B} \rangle
+|\mathbf{k},\mathbf{q},\mathsf{C} \rangle 
\right)
\label{eq:kq1}
\end{equation}
makes an SU(3) singlet, since all three fermions have been moved from $\mathbf{q}$ to $\mathbf{k + q}$, where they are anti-symmetrized. 
The linear combination orthogonal to $|\mathbf{k},\mathbf{q},\mathbf{1} \rangle$ belongs to the adjoint irreducible representation of SU(3), the $\mathbf{8}$. This is because the irreducible representation of a hole is $\mathbf{\bar 3}$ (the two remaining  fermions anti-symmetrize), the irreducible representation of the particle (a single fermion) is $\mathbf{3}$, and the product of these two representations is $\mathbf{3} \otimes \mathbf{\bar 3} = \mathbf{1} \oplus \mathbf{8}$. Fig.~\ref{fig:weight_diagrams} shows the weight diagrams of these irreducible representations. The combination in Eq.~(\ref{eq:kq1}) is the singlet $\mathbf{1}$, and the two orthogonal ones are the two states in the middle of the weight diagram of $\mathbf{8}$. In fact, 
the eight states $T^{a}_{\mathbf{k}} P_{\text{G}} | \text{FS} \rangle = \frac{1}{\sqrt{L}} \sum_{\mathbf{R}} e^{i \mathbf{k} \cdot \mathbf{R}} T^{a}_{\mathbf{R}} P_{\text{G}} | \text{FS} \rangle $ for $a \in \lbrace 1 \ldots 8 \rbrace$ form a basis for the adjoint representation $\mathbf{8}$, where $T^{a}_{\mathbf{R}}$ is given by Eq.~(\ref{eq:TGellMann}):
\begin{align}
\label{eq:TaPGFS}
    T^{a}_{\mathbf{k}} P_{\text{G}} | \text{FS} \rangle 
    &=  
    \frac{1}{\sqrt{L}} \sum_{\mathbf{R}} e^{i \mathbf{k} \cdot \mathbf{R}} T^{a}_{\mathbf{R}} P_{\text{G}} | \text{FS} \rangle  \nonumber\\
    &= \frac{1}{\sqrt{L}} \sum_{\mathbf{q}} \sum_{\sigma\sigma'} \frac{1}{2} P_{\text{G}}
    f^{\dagger}_{\mathbf{k} + \mathbf{q}, \mathsf{\sigma}} \lambda^a_{\sigma,\sigma'}f^{\phantom{\dagger}}_{\mathbf{q}, \mathsf{\sigma'}} P_{\text{G}}
    | \text{FS} \rangle \;, 
 \end{align}
where we used that the $T^{a}_{\mathbf{R}}$ and $P_{\text{G}}$ commute [see Eq.~(\ref{eq:comm_T_n})], and the Fourier transform convention
\begin{equation}
\label{Fourier transform convention of the annihilation operator}
    f^{\phantom{\dagger}}_{\mathbf{R}, \sigma} = \frac{1}{\sqrt{L}} \sum_{\mathbf{q}} e^{i\mathbf{q} \cdot \mathbf{R} } f^{\phantom{\dagger}}_{\mathbf{q}, \sigma} .
\end{equation}
In particular, applying the diagonal $T^3_{\mathbf{k}}$ and $T^8_{\mathbf{k}}$,  we get
\begin{subequations}
\begin{align}
\label{particle-hole excitations belong to the adjoint irrep}
    T^{3}_{\mathbf{k}} P_{\text{G}} | \text{FS} \rangle 
    &= 
    \frac{1}{\sqrt{L}} \sum_{\mathbf{q}} \frac{1}{2} P_{\text{G}}
    \left( 
    f^{\dagger}_{\mathbf{k} + \mathbf{q}, \mathsf{A}} f^{\phantom{\dagger}}_{\mathbf{q}, \mathsf{A}} - f^{\dagger}_{\mathbf{k} + \mathbf{q}, \mathsf{B}} f^{\phantom{\dagger}}_{\mathbf{q}, \mathsf{B}} 
    \right)
    | \text{FS} \rangle \nonumber \\
    &= \frac{1}{\sqrt{L}} \sum_{\mathbf{q}} \frac{1}{2} \left( |\mathbf{k}, \mathbf{q}, \mathsf{A} \rangle - |\mathbf{k}, \mathbf{q}, \mathsf{B} \rangle \right) \end{align}
and
\begin{equation}
\label{T^8 P_G |FS> belongs to the adjoint irrep}
    T^{8}_{\mathbf{k}} P_{\text{G}} | \text{FS} \rangle = \frac{1}{\sqrt{L}} \sum_{\mathbf{q}} \frac{1}{2 \sqrt{3}} \left( |\mathbf{k}, \mathbf{q}, \mathsf{A} \rangle + |\mathbf{k}, \mathbf{q}, \mathsf{B} \rangle - 2 |\mathbf{k}, \mathbf{q}, \mathsf{C} \rangle \right).
\end{equation}
\end{subequations}
Consequently, the linear combinations in the sums belong to the $\mathbf{8}$, and they are orthogonal to $|\mathbf{k},\mathbf{q},\mathbf{1} \rangle$, Eq.~(\ref{eq:kq1}).

\begin{table*}[btp]
\caption{Exact diagonalization (ED) and variational (vari.) results for chains of different length $L$. The ground state energy $E_{0}$, the gap $\Delta(k_{\text{min}})$ at the smallest momentum $k_{\text{min}}=2\pi/L$, the gap $\Delta(2\pi/3)$ and the weight of the lowest peak $S^{(0,0)}$ at $k=2\pi/3$, and the velocity $v = \Delta(k_{\text{min}}) / k_{\text{min}}$ are shown. }
\label{tab:ED_vari}
\begin{center}
\begin{ruledtabular}
\begin{tabular}{rcccccccccc}
  L & \multicolumn{2}{c}{$E_{0}$} & \multicolumn{2}{c}{$\Delta(k_{\text{min}})$} & \multicolumn{2}{c}{$\Delta(2\pi/3)$} & \multicolumn{2}{c}{$S^{(0,0)}$} & \multicolumn{2}{c}{$v$} \\
    & ED & vari. & ED & vari. & ED & vari. & ED & vari. & ED & vari.  \\
\hline
 9 & -4.78972 	& -4.78215 & 0.715188	& 0.721067 & 0.466118	& 0.464023 & 0.191144	& 0.192129 & 1.02443 & 1.03285 \\
12 & -6.31226	& -6.30017 & 0.545365	& 0.552532 & 0.350153	& 0.347568 & 0.178778	& 0.180269 & 1.04157 & 1.05526 \\
15 & -7.84810	& -7.83172 & 0.439298	& 0.446575 & 0.280624	& 0.277736 & 0.169014	& 0.170915 & 1.04875 & 1.06612 \\
18 & -9.39042	& -9.36986 & 0.367282	& 0.374293 & 0.234219	& 0.231136 & 0.161100	& 0.163321 & 1.05219 & 1.07227 \\ 
21 &-10.93635	&-10.91173 & 0.315346	& 0.321977 & 0.201024	& 0.197812 & 0.154520	& 0.157002 & 1.05396 & 1.07613 \\
\end{tabular}
\end{ruledtabular}
\end{center}
\label{default}
\end{table*}%

In $S^{33}(k,\omega)$ and $S^{88}(k,\omega)$ the relevant excited states $| \lambda \rangle$ are the ones which have non-zero overlap with $T^{3}_{\mathbf{k}}| 0 \rangle \approx T^{3}_{\mathbf{k}} P_{\text{G}}| \text{FS} \rangle$ and $T^{8}_{\mathbf{k}}| 0 \rangle \approx T^{8}_{\mathbf{k}} P_{\text{G}}| \text{FS} \rangle$, respectively. These states belong to the adjoint representation, therefore we have to look for the excited states $| \lambda \rangle$ in the subspace of states belonging to the same irreducible representation.  This explains why the particle-hole states (\ref{particle-hole excited states}) are useful in the description of the dynamical structure factor (\ref{dynamical structure factor}) of the kind $S^{33}(k,\omega)$ and $S^{88}(k,\omega)$, since their linear combinations belong to the adjoint representation as well. 

In the more general case of the SU($N$) symmetrical Heisenberg model the particle-hole excitations transform as $\mathbf{N} \otimes \overline{\mathbf{N}} = \mathbf{1} \oplus (\mathbf{N^2-1})$, so that their linear combinations will belong to the singlet and the $(N^2-1)$-dimensional adjoint representation -- the latter we get by acting with
the generators of the su($N$) algebra on the singlet ground state. Thus the particle-hole excitations (\ref{particle-hole excited states}) are useful for the calculation of the dynamical structure factor of the SU($N$) symmetric Heisenberg model for any $N$.

For the calculation of $S^{33}$ we can restrict ourselves to the subspace of states 
\begin{equation}
  |\mathbf{k},\mathbf{q},\mathbf{8}_3 \rangle = \frac{1}{2} \left( |\mathbf{k}, \mathbf{q}, \mathsf{A} \rangle - |\mathbf{k}, \mathbf{q}, \mathsf{B} \rangle \right) \;,
  \label{eq:kq83}
\end{equation} 
and similarly for $S^{88}$.

For later convenience, we use the states introduced by Ferrari {\it et al.} \cite{Becca}, 
\begin{equation}
\label{excited states of Becca}
|\mathbf{k},\mathbf{R},\sigma \rangle = P_{\text{G}} \frac{1}{\sqrt{L}} \sum_{\mathbf{R'}} e^{i\mathbf{k} \cdot \mathbf{R'}}f^{{\dagger}}_{\mathbf{R+R'},\sigma} f^{\phantom{\dagger}}_{\mathbf{R'},\sigma} |\text{FS} \rangle.
\end{equation}
instead of the states (\ref{particle-hole excited states}).
These are, following the convention in Eq.~(\ref{Fourier transform convention of the annihilation operator}), the Fourier transforms of the particle-hole excited states (\ref{particle-hole excited states}), since
\begin{equation}
\label{relation between Dalla and Becca states}
|\mathbf{k},\mathbf{q},\sigma \rangle = \frac{1}{\sqrt{L}} \sum_{\mathbf{R}} e^{i (\mathbf{k} + \mathbf{q})\cdot \mathbf{R} } | \mathbf{k}, \mathbf{R}, \sigma \rangle  .
\end{equation}
Since the two sets of states (\ref{particle-hole excited states}) and (\ref{excited states of Becca}) are equivalent up to a Fourier transformation, the linear combinations
\begin{equation}
\label{eq:kR83}
    |\mathbf{k},\mathbf{R},\mathbf{8}_3 \rangle = \frac{1}{\sqrt{2}}\left(
    |\mathbf{k},\mathbf{R}, \mathsf{A} \rangle 
    - |\mathbf{k},\mathbf{R}, \mathsf{B} \rangle \right)\;
\end{equation}
and
\begin{equation}
\label{eq:kR88}
    |\mathbf{k},\mathbf{R},\mathbf{8}_8 \rangle  = \frac{1}{\sqrt{6}}\left(
    |\mathbf{k},\mathbf{R}, \mathsf{A} \rangle
    +|\mathbf{k},\mathbf{R}, \mathsf{B} \rangle
    - 2|\mathbf{k},\mathbf{R}, \mathsf{C} \rangle 
    \right)
\end{equation}
also belong to the adjoint representation $\mathbf{8}$. 

The states $|\mathbf{k},\mathbf{R},\sigma \rangle$ are eigenstates of the translation operator with wave vector $ \mathbf{k} + \mathbf{k_{\text{FS}}}$, where $\mathbf{k_{\text{FS}}}$ is the wave vector of the Fermi sea. Therefore, the projected Hamiltonian 
\begin{equation}
    \tilde{\mathcal{H}}^{\mathbf{k}}_{\mathbf{R}, \sigma ; \mathbf{R'},\sigma'} = \langle \mathbf{k} , \mathbf{R}, \sigma | \mathcal{H} | \mathbf{k} , \mathbf{R'}, \sigma' \rangle,
\label{eq:HRR}
\end{equation}
and the overlap matrix
\begin{equation}
    \mathcal{O}^{\mathbf{k}}_{\mathbf{R}, \sigma;\mathbf{R'},\sigma'} = \langle \mathbf{k} , \mathbf{R}, \sigma |  \mathbf{k} , \mathbf{R'}, \sigma' \rangle ,
        \label{eq:ORR}
\end{equation}
are block diagonal in $\mathbf{k}$ for translationally invariant systems (like ours), where $\mathbf{R}, \sigma$ can be thought of as a row index and $\mathbf{R'},\sigma'$ as a column index. These matrices were evaluated by a Monte Carlo method described in Appendix \ref{sec:Monte_Carlo_of_H_and_O}.

In order to find the eigenstates of the projected Hamiltonian in this truncated Hilbert space, we need to solve the generalized eigenvalue problem for the block matrices 
\begin{equation}
\label{generalized eigenvalue problem}
   \tilde{\mathcal{H}}^{\mathbf{k}} | \phi^{\mathbf{k}}_n \rangle = E^{\mathbf{k}}_n \mathcal{O}^{\mathbf{k}} | \phi^{\mathbf{k}}_n \rangle
\end{equation}
(details of the generalized eigenvalue problem are given in Appendix~\ref{sec:gen_eigenvalue}). Then, the eigenstates of the blocks of the projected Hamiltonian $\tilde{\mathcal{H}}^{\mathbf{k}}$
are
\begin{equation}
\label{eigenstates of the projected Hamiltonian}
    | \phi^{\mathbf{k}}_n \rangle = \sum_{\mathbf{R}, \sigma} A^{n, \mathbf{k}}_{\mathbf{R}, \sigma} | \mathbf{k}, \mathbf{R}, \sigma \rangle.
\end{equation}
Following Ferrari {\it et al.} \cite{Becca}, we can write:
\begin{align}
    T^{3}_{\mathbf{k}} P_{\text{G}} | \text{FS} \rangle 
    &=  \frac{1}{\sqrt{L}} \sum_{\mathbf{R}} e^{i \mathbf{k} \cdot \mathbf{R}}  T^{3}_{\mathbf{R}} P_{\text{G}} | \text{FS} \rangle  \\
    &=  \frac{1}{\sqrt{L}} \sum_{\mathbf{R}} e^{i \mathbf{k} \cdot \mathbf{R}} P_{\text{G}} \frac{1}{2}
    \left( 
    f^{\dagger}_{\mathbf{R}, \mathsf{A}} f^{\phantom{\dagger}}_{\mathbf{R}, \mathsf{A}} - f^{\dagger}_{\mathbf{R}, \mathsf{B}} f^{\phantom{\dagger}}_{\mathbf{R}, \mathsf{B}} \right)  | \text{FS} \rangle \nonumber \\
    &= \frac{1}{2} \left( |\mathbf{k}, 0, \mathsf{A} \rangle - |\mathbf{k}, 0, \mathsf{B} \rangle \right), \nonumber
\end{align}
where we used again that $[ T^{3}_{\mathbf{R}},P_{\text{G}}]=0$, see Eq.~(\ref{eq:comm_T_n}). Consequently,  the matrix elements for $S^{33}(\mathbf{k},\omega)$ are 
\begin{align}
\label{sandwich of T3 expressed with the overlap matrix}
    \langle \phi^{\mathbf{k}}_n  
    | T^{3}_{\mathbf{k}} P_{\text{G}}|\text{FS} \rangle 
    &=\frac{1}{2}\left( \langle \phi^{\mathbf{k}}_n  |\mathbf{k}, 0, \mathsf{A} \rangle - \langle \phi^{\mathbf{k}}_n  |\mathbf{k}, 0, \mathsf{B} \rangle \right) \\
    &= \frac{1}{2} \sum_{\mathbf{R},\sigma} (A^{n,\mathbf{k}}_{\mathbf{R},\sigma})^* (\mathcal{O}_{\mathbf{R}, \sigma ; 0,\mathsf{A}}  -  \mathcal{O}_{\mathbf{R}, \sigma ; 0,\mathsf{B}} ), \nonumber
\end{align}
In order to get the correct weights for $S^{33}(\mathbf{k},\omega)$,
 we normalize it so that the sum rule $\frac{1}{L}\sum_{\mathbf{k}}\int d\omega S^{33}(\mathbf{k},\omega) = \frac{1}{6}$ [Eq.~(\ref{eq:S33ksumrule})] is satisfied. The normalization is needed, because the approximating ground state $P_{\text{G}} |\text{FS} \rangle$ is not normalized. Analogous equations hold for the matrix elements in $ S^{88}(\mathbf{k},\omega) $
 %The reason for this is that the Fermi sea $|\text{FS} \rangle$ is supposed to be normalized, and the action of the Gutzwiller projector $P_{\text{G}} | \text{FS} \rangle$ decreases its norm.

There is an alternative route to calculate the structure factor and to fulfill the sum rule, following Li and Yang \cite{First_S_q_w_latter,*Yang_Li_2011PhRvB..83f4524Y}. Instead of replacing the exact ground state in the expression $\langle \lambda | T^{a}_{\mathbf{k}}| 0 \rangle $ by our approximating ground state, we can replace it with the lowest energy eigenstate of the projected Hamiltonian $\tilde{\mathcal{H}}^{\mathbf{0}}$ in the $\mathbf{k} = \mathbf{0} $ singlet sector. This state is already normalized with respect to the overlap matrix, and it may even have lower energy than the approximating ground state we have started with. However, it turns out that in our case the only linearly independent state between the excitations $| \mathbf{k} = \mathbf{0}, \mathbf{R}, \sigma \rangle$ is the approximating ground state we have started with, therefore mixing the approximating ground state with the excited states will not yield a better ground state, and the only effect of this procedure is the normalization of $P_{\text{G}}|\text{FS} \rangle$. This method automatically fulfills the sum rules without any statistical error, but it gives just the same result for $S^{33}(\mathbf{k},\omega)$ as the method of Ferrari {\it et al.}  presented above\cite{Becca}, after enforcing the sum rule. On the other hand, in order to calculate the term $\langle \lambda | T^{3}_{\mathbf{k}}| 0 \rangle $, the method of  Li and Yang  requires in addition the measurement of $\langle \mathbf{k}, \mathbf{R}, \sigma | T^{3}_{\mathbf{k}} | \mathbf{0}, \mathbf{R'}, \sigma' \rangle$ \cite{First_S_q_w_latter,*Yang_Li_2011PhRvB..83f4524Y}, which is not needed for the method of Ferrari {\it et al.} \cite{Becca}.

  In Tab.~\ref{tab:ED_vari} we compare the ED and the variational method (taking into account all single particle-hole excitations) for small system sizes (up to $L=21$). For larger systems, we need to apply Monte Carlo sampling. This technical part is described in Appendix~\ref{sec:Monte_Carlo_of_H_and_O}.

%%%%%%%%%%%%%%%%%%%%%%%%%%%%%%%%%%%%%%%%%%%%%%%%%%%%%%%
\subsection{Results}
%%%%%%%%%%%%%%%%%%%%%%%%%%%%%%%%%%%%%%%%%%%%%%%%%%%%%%%

%%%%%%%%%%%%%%%%%%%%%%%%%%%%%%%%%%%%%%%%%%%%%%%%%%%%%%%
\begin{figure}[b]
	\centering
\includegraphics[width=0.95\columnwidth]{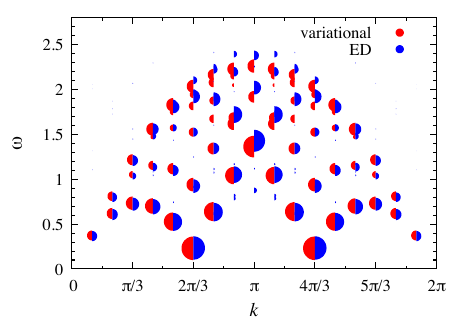}
	\caption{ 
	Comparing $S^{33}(k,\omega)$ for $L=18$ calculated using Gutzwiller projected one particle-hole excitations (red) and by ED (blue). The area of the circles is proportional to the matrix element squared.}
		 \label{fig:Skw_L18}
\end{figure}
%%%%%%%%%%%%%%%%%%%%%%%%%%%%%%%%%%%%%%%%%%%%%%%%%%%%%%%

%%%%%%%%%%%%%%%%%%%%%%%%%%%%%%%%%%%%%%%%%%%%%%%%%%%%%%%
\begin{figure}[t]
	\centering
\includegraphics[width=0.95\columnwidth]{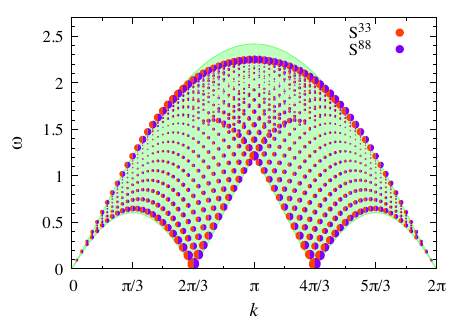}
	\caption{ 
	 $S^{33}(k,\omega)$ and $S^{88}(k,\omega)$ for $L=72$ calculated by VMC. The area of the circles is proportional to the matrix element squared. The green background shows the  two soliton continuum $(k,\omega)=(q_{\mathbf{3}\mathbf{\bar 3}},
  \varepsilon_{\mathbf{3}\mathbf{\bar 3}})$ of the Bethe Ansatz solution, 
  Eqs.~(\ref{eq:solitons}) and (\ref{eq:2solcont}), in the thermodynamic limit. For this plot, the total number of uncorrelated measurements was $10^8$.
}
	 \label{fig:Skw_largesystem}
\end{figure}
%%%%%%%%%%%%%%%%%%%%%%%%%%%%%%%%%%%%%%%%%%%%%%%%%%%%%%%

First, we calculated the $S^{33}(k,\omega)$ for a small ($L=18$) system by exact evaluation of the Hamiltonian and overlap matrices, $\tilde{\mathcal{H}}^{\mathbf{k}}_{\mathbf{R}, \sigma;\mathbf{R'},\sigma'}$ and $\mathcal{O}^{\mathbf{k}}_{\mathbf{R}, \sigma;\mathbf{R'},\sigma'}$, by summing over all the possible $|x\rangle$ states in Eqs.~(\ref{eq:HOx}). The result is shown in Fig.~\ref{fig:Skw_L18}, together with the dynamical structure factor calculated by exact diagonalization (ED), with the help of the standard Lánczos algorithm \cite{1987PhRvL..59.2999G}. We also calculated $S^{88}(k,\omega)$ in order to compare it to $S^{33}(k,\omega)$, and as expected, the two structure factors were in perfect correspondence.

Next, using the importance sampling introduced in Eq.~(\ref{max of abs of g matrix}) of Appendix \ref{sec:Monte_Carlo_of_H_and_O}, we performed a Monte Carlo evaluation of the Hamiltonian and overlap matrices in the reduced Hilbert space for $L=72$. The result is shown in Fig.~\ref{fig:Skw_largesystem} for both $S^{33}(k,\omega)$ and $S^{88}(k,\omega)$, which are indistinguishable in the figure.  This can be compared to the dynamical structure factor calculated by the matrix product state  (MPS) algorithm with infinite boundary conditions (the $\theta=\pi/4$ panel of Fig.~3 in \cite{DMRG_Binder_PRB2020}). 

  Careful examination of the results in Figs.~\ref{fig:Skw_L18} and \ref{fig:Skw_largesystem} reveals that the main features of the spectra are well reproduced, specifically the continuum and the disappearance of the gap at $k=2\pi/3$ and $4\pi/3$. The discrepancies from the exact result are negligible at low energies. At higher energies, above $\omega \gtrsim J$, the weights are shifted by about 10\% in energy. In Fig.~\ref{fig:Skw_L18} we also see that the lowest energy weights connecting the two towers at $k=2\pi/3$ and $k=4\pi/3$ 
are also missing. These are 4-soliton excitations, which are not captured by the 1 particle-hole Ansatz we use. The absence of the 4-soliton excitations is also obvious for the L=72 site result, when comparing to the MPS result \cite{DMRG_Binder_PRB2020}. 
 
 The elementary excitations from the Bethe-Ansatz solution are solitons with dispersion
\begin{subequations}
\label{eq:solitons}
\begin{align}
\varepsilon_{\mathbf{\bar 3}} (k) &= \frac{2 \pi }{3 \sqrt{3}} \left[\cos \left(\frac{\pi }{3}-k\right)-\cos \frac{\pi }{3} \right],&  0\leq k\leq \frac{2 \pi }{3} \;,
\\
\varepsilon_{\mathbf{3}}(k) &= \frac{2 \pi }{3 \sqrt{3}} \left[\cos \frac{\pi }{3} -\cos \left(k+\frac{\pi }{3}\right)\right], &  0\leq k\leq \frac{4 \pi }{3} \;.
\end{align}
\end{subequations}
in the thermodynamic limit \cite{1975PhRvB..12.3795S}.
The two-soliton continuum is spanned by a $\mathbf{\bar 3}$ and a $\mathbf{3}$ soliton, defined by
\begin{subequations}
\label{eq:2solcont}
\begin{align}
  q_{\mathbf{3}\mathbf{\bar 3}} & = k_{\mathbf{3}} +  k_{\mathbf{\bar 3}}\;, \\
  \varepsilon_{\mathbf{3}\mathbf{\bar 3}} & = \varepsilon_{\mathbf{3}} (k_{\mathbf{3}})  + \varepsilon_{\mathbf{\bar 3}} (k_{\mathbf{\bar 3}}) \;,
\end{align} 
\end{subequations}
where $k_{\mathbf{3}} \in [0,2\pi/3]$ and  $k_{\mathbf{\bar 3}} \in [0,4\pi/3]$. The main contribution to the dynamical structure factor comes from these two-soliton excitations, highlighted by green in Fig.~\ref{fig:Skw_largesystem}. Again, the agreement is remarkable, only at the higher energies around $k=\pi$ there is a noticeable discrepancy. 

  The solitons correspond to the excitations shown in Fig.~\ref{fig:particle-hole excitation}: the particles match with the $\varepsilon_{\mathbf{3}}(k)$ solitons, and the holes are the analogs of the $\varepsilon_{\mathbf{\bar 3}}(k)$. In the case of the SU(3) Haldane-Shastry model, the corresponding excitations, named  colorons, were considered in Refs.~\cite{2005EL.....71..987S,2006PhRvB..73w5105S}.
  
\subsubsection{The low-energy structure of a tower}

%%%%%%%%%%%%%%%%%%%%%%%%%%%%%%%%%%%%%%%%%%%%%%%%%%%%%%%
\begin{figure}[bt]
\centering
\includegraphics[width=0.75\columnwidth]{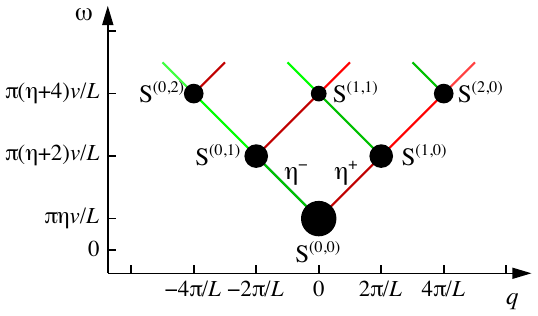}
	\caption{ 
 The finite-size structure of a tower at low energies. $S^{(0,0)}$ denotes the weight of the lowest energy peak in the tower, with energy $E_{0,0}$ and momentum $k_{0,0}$, the $S^{(i,i')}$ are the peaks in the tower following the notation in Eqs.~(\ref{eq:tower_eiikii}). The momenta are measured from $k_0$, the momentum of the lowest energy peak in the tower. We expect the ratio between the weights of $S^{(1,0)}$ and $S^{(0,0)}$ peaks to give the exponent $\eta^+$, similarly $\eta^- = S^{(0,1)}/S^{(0,0)}$.}
	\label{fig:tower_structure}
\end{figure}
%%%%%%%%%%%%%%%%%%%%%%%%%%%%%%%%%%%%%%%%%%%%%%%%%%%%%%%

%%%%%%%%%%%%%%%%%%%%%%%%%%%%%%%%%%%%%%%%%%%%%%%%%%%%%%%
\begin{figure}[t]
	\centering
	    \includegraphics[width=0.85\columnwidth]{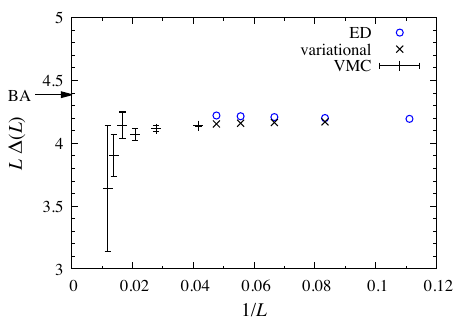}
	\caption{ 
	Finite size gap at the $k=2\pi/3$ multiplied by the system size L, as a function of $1/L$. The blue circles show the ED gap, the VMC results are shown by black crosses. The arrow points to $\pi \eta v_{\text{BA}} = 4 \pi^2/9$, the known value in the thermodynamic limit using $\eta =4/3$ and $v_{\text{BA}}= \pi/3$, the velocity from the Bethe Ansatz, Eq.~(\ref{eq:vBA}).
		}
	\label{fig:gap2pio3_finite_size}
\end{figure}
%%%%%%%%%%%%%%%%%%%%%%%%%%%%%%%%%%%%%%%%%%%%%%%%%%%%%%%

Does the  overall remarkable agreement also hold for the detailed low energy properties of the tower of excitations at $k=2\pi/3$ and $k=4\pi/3$? According to the conformal field theory, the energy and momenta of the excitations in a tower (see Fig.~\ref{fig:tower_structure}) are defined by
\begin{subequations}
\label{eq:tower_eiikii}
 \begin{align}
   E_{i,i'} - E_{0} &= \frac{\pi}{L} v (\eta^+ + \eta^-) + \frac{2\pi}{L} v (i+i') \;,\\ 
   k_{i,i'} - k_{0} &= \frac{\pi}{L} (\eta^+ - \eta^-) + \frac{2\pi}{L} (i - i') \;,
 \end{align}
\end{subequations}
where $\eta^+ + \eta^- = \eta$ in Eq.~(\ref{power law behaviour of the static peak}). 
The finite-size corrections of the energy gap between the bottom of the tower at $k=2\pi/3$ and the ground state energy should scale as 
\begin{equation}
\Delta(L) = E_{0,0} - E_{0} = \frac{\pi}{L} v \eta \;,
\label{eq:DeltaL}
\end{equation}
where $\eta$ =4/3 is the static exponent. Here we neglect logarithmic corrections\cite{1997PhRvB..55.8295I}. To verify the above formula, we  plot the $L \Delta(L)$ in Fig.~\ref{fig:gap2pio3_finite_size} from the different methods. We find that the $L\to\infty$ value is accurate to   about 10\%.

In Ref.~\cite{1997PhRvB..5515475P} the following relation has been found for the peaks of a tower originating from overlap determinants in the thermodynamic limit :
 \begin{equation}
   \frac{S^{(i,i')}}{S^{(0,0)}} = \frac{\Gamma( i + \eta^+ )}{\Gamma(i+1) \Gamma(\eta^+)} \frac{\Gamma( i' + \eta^- )} {\Gamma(i'+1) \Gamma(\eta^-)} 
   \label{eq:tower_sii}
 \end{equation}
Combining the asymptotic expansion of the $\Gamma$-functions
\begin{equation}
  \frac{\Gamma(i+\eta)}{\Gamma(i+1)} \approx \left(i+\frac{\eta}{2} \right)^{\eta-1}
\end{equation}
with the finite-size expressions for the energy and momenta, Eqs.~(\ref{eq:tower_eiikii}), 
we get the expected power-law behavior of the dynamical correlation function 
\begin{align}
 S(k,\omega) &\propto \sum_{i,i'} S^{(i,i')} \delta(k-k_{i,i'}) \delta(\omega - E_{i,i'} +E_{0}) \nonumber
 \\
 &\propto  \begin{cases} 
       (\omega + v q )^{\eta^+ - 1} (\omega - v q )^{\eta^- - 1} \;,& \omega \geq v|q| \,; \\
       0 \;, & \omega< v|q| 
   \end{cases}
\end{align}
for $L\to\infty$, where $q = k-k_0$ is the relative momentum. Integrating over $\omega$, we recover the power-law singularity of the structure factor,  
\begin{equation}
  S(k) \propto q^{\eta^+ + \eta^- - 1} = q^{\eta - 1}, 
\end{equation}
see Eq.~(\ref{power law behaviour of the static peak}).

\begin{figure}[t]
\centering
\includegraphics[width=0.9\columnwidth]{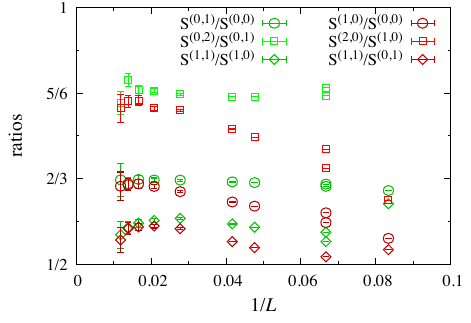}
	\caption{ 
	The ratios of the weights of the low energy peaks. The ratios provide information about the exponents, see Eqs.~(\ref{eq:ratios}) and (\ref{eq:ratios2}).}
	\label{fig:ratios}
\end{figure}

Assuming that Eq.~(\ref{eq:tower_sii}) holds more generally, we can get the exponents from the ratios of the lowest lying weights as
\begin{equation}
  \frac{S^{(1,0)}}{S^{(0,0)}} = \eta^+  \quad\text{and}\quad \frac{S^{(0,1)}}{S^{(0,0)}} = \eta^-  \;.
      \label{eq:ratios}
\end{equation}
The ratios between the higher lying weights 
\begin{equation}
    \frac{S^{(2,0)}}{S^{(1,0)}} = \frac{1+\eta^+}{2} \quad\text{and}\quad \frac{S^{(0,2)}}{S^{(0,1)}} = \frac{1+\eta^-}{2} 
    \label{eq:ratios2}
\end{equation}
may serve to check the validity of the assumption. In Fig.~\ref{fig:ratios} we plot the ratios for different system sizes. It appears that both $S^{(1,0)}/S^{(0,0)}$ and $S^{(0,1)}/S^{(0,0)}$ tend to the exponents $\eta^+ = \eta^- = 2/3$ (so that $\eta^+ + \eta^- = \eta = 4/3$). The ratios $S^{(2,0)}/S^{(1,0)}$ and $S^{(0,2)}/S^{(0,1)}$ go to 5/6, which is in accordance with Eq.~(\ref{eq:ratios2}). The ratios including the higher lying $(1,1)$ peak -- $S^{(1,1)}/S^{(1,0)}$ and $S^{(1,1)}S^{(0,1)}$
 -- are less conclusive, they are more scattered (these ratios should also go to $2/3$).
 
\begin{figure}[t]
\centering
\includegraphics[width=0.9\columnwidth]{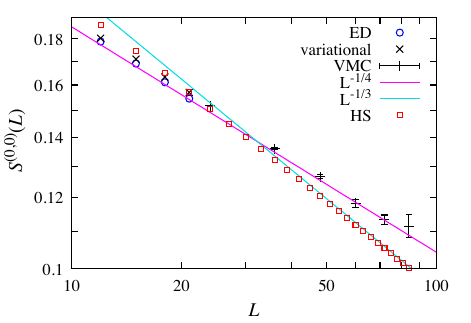}
	\caption{ 
	Scaling of the lowest peak from ED, variational, and VMC methods for the Heisenberg model and the exact values for the Haldane-Shastry model, together with the asymptotic $L^{-1/3}$ power-law behavior (cyan line). The $L^{-1/4}$ magenta line is a guide to the eye. Note the slight downward bending of the ED data which suggests that the exponent is in fact smaller than $-1/4$, tending toward $-1/3$.  }
	\label{fig:mex_scaling}
\end{figure}

 Fig.~\ref{fig:mex_scaling} shows the scaling of the weight at the bottom of the tower in a log-log plot. It shall go as
\begin{equation}
S^{(0,0)} \propto L^{1-\eta}
\end{equation}
with the system size. It is hard to get a definite value for the exponent, but  $1-\eta$ appears to be closer to -0.25 than to -1/3. The smaller exponent would also explain the finite-size scaling of the gap, shown in Fig.~\ref{fig:gap2pio3_finite_size}, as $\pi \eta v_{\text{BA}}\approx 4.11$ with the $1-\eta = -0.25$. However, an exponent different from $-1/3$ would make it difficult to explain the almost perfect $-1/3$ exponent in the non-analytical part of the $S^{33}(k)$, as it likely originates from the tower. 

%%%%%%%%%%%%%%%%%%%%%%%%%%%%%%%%%%%%%%%%%%%%%%%%%%%%%%%
%%%%%%%%%%%%%%%%%%%%%%%%%%%%%%%%%%%%%%%%%%%%%%%%%%%%%%%

%%%%%%%%%%%%%%%%%%%%%%%%%%%%%%%%%%%%%%%%%%%%%%%%%%%%%%%
%%%%%%%%%%%%%%%%%%%%%%%%%%%%%%%%%%%%%%%%%%%%%%%%%%%%%%%

%%%%%%%%%%%%%%%%%%%%%%%%%%%%%%%%%%%%%%%%%%%%%%%%%%%%%%%
\begin{figure}[t]
\centering
\includegraphics[width=0.7\columnwidth]{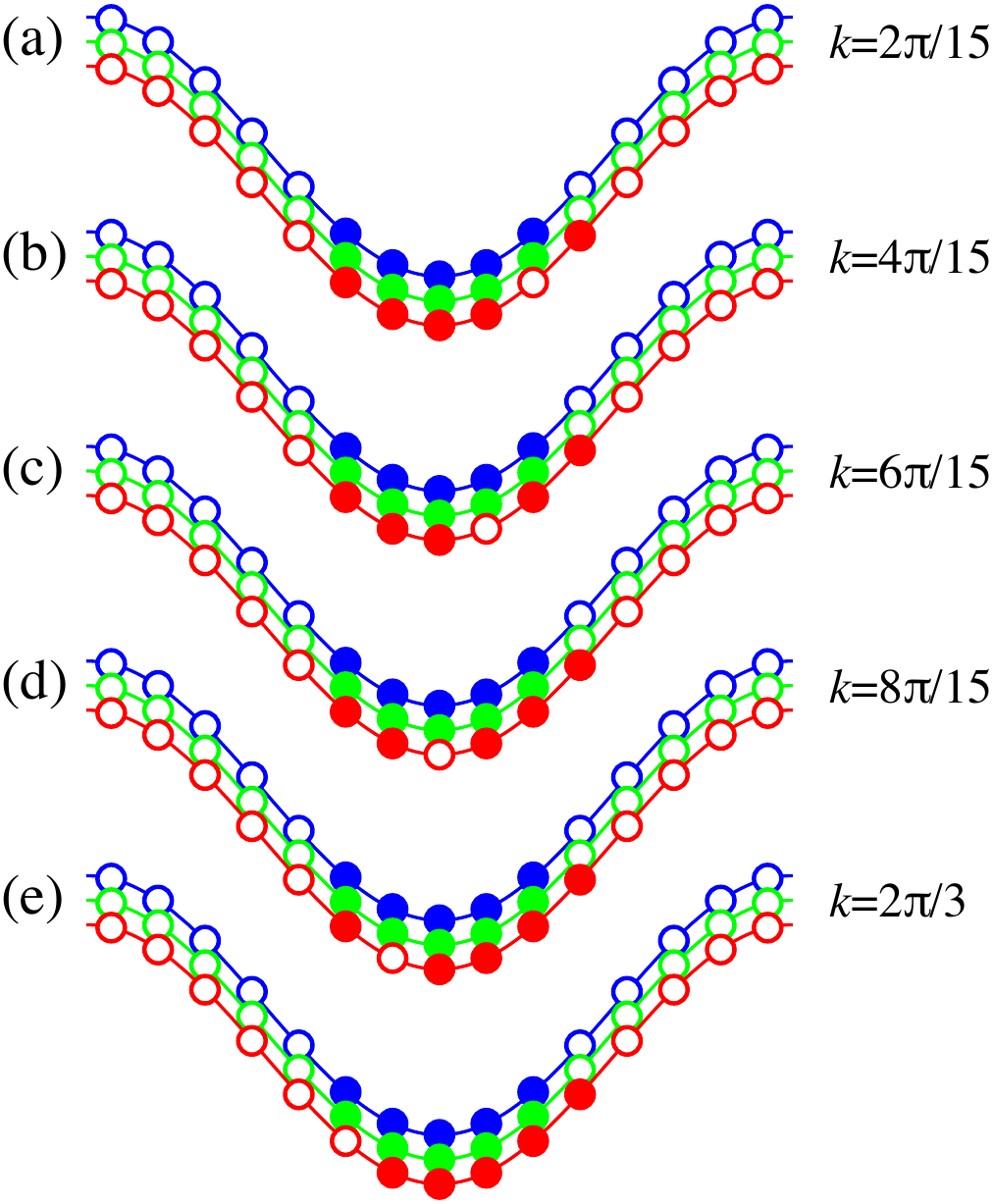}
	\caption{ 
	The configurations forming the `arc' -- the lowest energy excitations for momenta from $k=2\pi/L$ (a) to $k=2\pi/3$ (e) -- in the Haldane-Shastry model for $L=15$. The Gutzwiller projected wave function of these particle-hole excitations are exact eigenstates of the Hamiltonian Eq.~(\ref{eq:HamiltonHSSU3}).}
	\label{fig:arc_HS}
\end{figure}
%%%%%%%%%%%%%%%%%%%%%%%%%%%%%%%%%%%%%%%%%%%%%%%%%%%%%%%

\subsubsection{Haldane-Shastry model}
 
 We applied the variational method with exact evaluation of Eqs.~(\ref{eq:HOx}) for small system sizes ($L=9,12$ and 15). Solving the generalized eigenvalue problem, the finite-size gap at $k=2\pi/3$ is precisely equal to
 \begin{equation}
  \Delta_{2\pi/3} = \frac{2}{3} \frac{\pi^2}{L} = \frac{\pi}{L} v_{\text{HS}} \frac{4}{3} 
 \end{equation}
 for all the system sizes we considered. Comparing with Eq.~(\ref{eq:DeltaL}), we can read off the exponent $\eta = 4/3$, which is the same as the one of the Heisenberg model with nearest neighbor exchange only. 
 
 We also calculated the dynamical structure factor for L=15 variationally by taking into account the one particle-hole excitations, the energies and the weights for different momenta are presented in Tab. II. The exact analytical form of the dynamical structure factor for the SU(2) symmetric Haldane-Shastry model was determined in Refs. \cite{LESAGE1995585, *PhysRevLett.73.1574} and for the SU(N) model in Refs. \cite{2000PhRvL..84.1308Y} and \cite{2000JPSJ...69..900Y}. We compared our results to Table I in Ref. \cite{2000JPSJ...69..900Y}.
 We also calculated the dynamical structure factor for $L=15$ and compared it to Table 1 in \cite{2000JPSJ...69..900Y}, where the exact analytical result is given (see also \cite{2000PhRvL..84.1308Y}).  It turns out that at smaller momenta our variational treatment gives the correct excited states of the Haldane-Shastry model, including the bottom of the two towers at $k=2\pi/3$ and $k=4\pi/3$. In particular, the energies in units of $(\pi/L)^2$ are all integers for the exact eigenstates, as noted in Ref.~\cite{2000JPSJ...69..900Y} (see also Ref.~\cite{1988PhRvL..60..635H} for the SU(2) model).
In addition, there are some peaks for which the energy is exact, but the weight is smaller. The explanation is that the Haldane-Shastry model possesses a higher, Yangian symmetry, and the one particle-hole states are degenerate with other states not described by the variational Ansatz. Further investigations of small systems (up to $L=15$) revealed that the (Gutzwiller projected) particle-hole excitations shown in Fig.~\ref{fig:arc_HS} are exact eigenstates of the Haldane-Shastry Hamiltonian and form the arc of the lowest  energy excitations from $k=2\pi/L$ (Fig.~\ref{fig:arc_HS}(a)) to $k=2\pi/3$ (Fig.~\ref{fig:arc_HS}(e)). They are analogous to the des Cloizeaux-Pearson branch in the SU(2) Heisenberg model.  For these states the $\mathbf{1}$ and $\mathbf{8}$ are degenerate, manifesting the higher Yangian symmetry of the model \cite{1992PhRvL..69.2021H}.  The detailed examination of the $L=6, 9, 12,$ and $15$ systems allowed for the extrapolation of the momenta, energies, and the weights in the dynamical structure factor for the states in the arc:
\begin{subequations}
\label{eq:arc}
\begin{align}
 k_j &= \frac{2\pi j }{L}\;,\\
  \omega_j &= \left(\frac{\pi}{L} \right)^2 j (L + 2 - 3 j) \;, \\
S^{33}_j &=  
\frac{1}{6}
\frac{\Gamma \left(\frac{2}{3}\right) \Gamma (j) \Gamma \left(\frac{L}{3}\right) \Gamma \left(\frac{L}{3}-j+\frac{2}{3}\right)}{\Gamma \left(j-\frac{1}{3}\right) \Gamma \left(\frac{L}{3}+\frac{2}{3}\right) \Gamma \left(\frac{L}{3}-j+1\right)}  \;,
\label{eq:S33j}
 \end{align}
\end{subequations}
where $1\leq j \leq L/3$. They coincides with the exact expressions for the corresponding excitations with quantum numbers $c_1 = j$ and $c_2=c_3=0$ presented in \cite{2000JPSJ...69..900Y}.

From these expressions, the weight of the bottom of the tower is $S^{(0,0)}(L) = S^{33}_{L/3}$,
\begin{align}
S^{(0,0)}(L) &=  
\frac{1}{6}
\frac{\Gamma^2 \left(\frac{2}{3}\right) \Gamma^2(\frac{L}{3}) }{\Gamma \left(\frac{L}{3}-\frac{1}{3}\right) \Gamma \left(\frac{L}{3}+\frac{2}{3}\right) } \;.
\end{align}
The asymptotic expansion in the $L \to \infty $ limit gives the
\begin{align}
S^{(0,0)}(L) &=  \frac{\Gamma \left(\frac{2}{3}\right)^2}{2\ 3^{2/3} }
L^{-\frac{1}{3}}\left(1 - \frac{1}{3L} + \cdots\right)
\end{align}
$L^{-1/3}$ power-law behavior, shown in Fig~\ref{fig:mex_scaling}.
The ratios in Eq.~(\ref{eq:ratios}) are also fulfilled,
\begin{align}
\frac{S^{(0,1)}(L)}{S^{(0,0)}(L)} = 
\frac{S^{33}_{L/3-1}}{S^{33}_{L/3}} 
&= \frac{2 (L-4)}{3 (L-3)} = \frac{2}{3} \left(1 - \frac{1}{L} + \cdots\right)\;,
\end{align}
with the exponent $\eta^- = 2/3$. In fact, replacing $j$ by $L/3-i'$ into Eq.~(\ref{eq:S33j}), where $i'$ measures the distance from the bottom of the tower at $k=2\pi/3$, and taking the $i' \ll L$ limit, we get 
%\begin{align}
%\frac{S^{(0,i')}(L)}{S^{(0,0)}(L)} &= 
%\frac{S^{33}_{L/3-i'}}{S^{33}_{L/3}} \nonumber\\
%%&= 
%%\frac{ \Gamma (\frac{L}{3}-i')  \Gamma \left(\frac{L}{3}-\frac{1}{3}\right)  
%%}{
%%\Gamma \left(\frac{L}{3}-i'-\frac{1}{3}\right)  \Gamma(\frac{L}{3})} 
%%\frac{\Gamma \left(i'+\frac{2}{3}\right)  
%%}{
%%  \Gamma \left(i'+1\right) \Gamma\left(\frac{2}{3}\right) } \;.
%%  \\
%&= 
%\frac{\Gamma \left(i'+\frac{2}{3}\right)  
%}{
%  \Gamma \left(i'+1\right) \Gamma\left(\frac{2}{3}\right) } \left(1 - \frac{i'}{L} +\cdots \right)\;.
%\end{align}
\begin{equation}
\frac{S^{(0,i')}(L)}{S^{(0,0)}(L)} = 
\frac{S^{33}_{L/3-i'}}{S^{33}_{L/3}} 
= 
\frac{\Gamma \left(i'+\frac{2}{3}\right)  
}{
  \Gamma \left(i'+1\right) \Gamma\left(\frac{2}{3}\right) } \left(1 - \frac{i'}{L} +\cdots \right)\;,
\end{equation}
just what we expect from Eq.~(\ref{eq:tower_sii}).

\begin{table}[htp]
\caption{The energies $\omega$ and the weights $S^{33}(k,\omega)$ in the dynamical structure factor of the Haldane-Shastry model, evaluated for the $L=15$ site chain using the Gutzwiller projected variational basis. To facilitate an easy comparison with the exact result presented in \cite{2000JPSJ...69..900Y}, we have multiplied our $S^{33}(k,\omega)$ data (column 4) by 2 in column 5. The `--' in column 5 means that this peak is not exact, the ` * ' that the energy is exact, but the weight is not exhausted by the single particle-hole excitations. The last column gives the correspondence to the momenta configurations shown in Fig.~\ref{fig:arc_HS} for states in the `arc',  described by Eqs.~(\ref{eq:arc}).}
\label{tab:HSVMC}
\begin{center}
\begin{ruledtabular}
\begin{tabular}{clcllc}
 $15 k /2\pi$ & $\omega/J$ & $225\omega/\pi^2$ & $S^{33}(k,\omega)$ & 2 $S^{33}(k,\omega)$ & arc \\
\hline
 1 & 0.614109  & 14 & 0.035714 & 0.071429 & (a) \\
 2 & 0.965028  & 22 & 0.058442 & 0.116883 & (b) \\
 2 & 1.140488  & 26 & 0.019231 & 0.038461 &   \\
 3 & 1.052758  & 24 & 0.078896 & 0.157792 & (c) \\
 3 & 1.491407  & 34 & 0.050350 & 0.100699 &   \\
 3 & 1.491407  & 34 & 0 & 0  &   \\
 4 & 0.877298  & 20 & 0.106510 & 0.213019 & (d) \\
 4 & 1.579137  & 36 & 0.049170 & 0.098339* &   \\
 4 & 1.684384  & -- & 0.021417 &  -- &   \\
 4 & 1.884737  & -- & 0.021572 &  -- &   \\
 5 & 0.438649  & 10 & 0.174289 & 0.348577  & (e) \\
 5 & 1.403677  & 32 & 0.069281 & 0.138562*  &   \\
 5 & 1.713835  & -- & 0.025216 &  -- &   \\
 5 & 1.941583  & -- & 0.025474 &  -- &   \\
 5 & 2.077312  & -- & 0.030178 &  -- &   \\
 6 & 0.965028  & 22 & 0.115975 & 0.231951*  &   \\
 6 & 1.543002  & -- & 0.028316 &   --&   \\
 6 & 1.770273  & -- & 0.048978 &  -- &   \\
 6 & 2.093077  & -- & 0.028767 &  -- &   \\
 6 & 2.201815  & -- & 0.030093 &  -- &   \\
 7 & 1.330560  & -- & 0.104132 &  -- &   \\
 7 & 1.540425  & -- & 0.044968 &  -- &   \\
 7 & 1.960484  & -- & 0.022560 &  -- &   \\
 7 & 2.058017  & -- & 0.009248 &  -- &   \\
 7 & 2.243803  & -- & 0.051222 &  -- &   \\
\end{tabular}
\end{ruledtabular}
\end{center}
\label{default}
\end{table}%

%%%%%%%%%%%%%%%%%%%%%%%%%%%%%%%%%%%%%%%%%%%%%%%%%%%%%%%
\section{Conclusion}
\label{sec:Conlusion}
%%%%%%%%%%%%%%%%%%%%%%%%%%%%%%%%%%%%%%%%%%%%%%%%%%%%%%%

To conclude, we extended the dynamical VMC method of \cite{First_S_q_w_latter,*Yang_Li_2011PhRvB..83f4524Y,Excited_states_above_Fermi_sea,Mei_Wen_arxiv_2015,Becca} to the case of the SU(3) Heisenberg model. To describe the correlated states of the SU(3) spins, we used the Gutzwiller projected Fermi sea of three-color fermions as a variational ground state and built the spectrum from single particle-hole excitations.

On the technical side, we modified the importance sampling used by Li and Yang in Ref.~\cite{First_S_q_w_latter,*Yang_Li_2011PhRvB..83f4524Y}: instead of selecting a configuration based on its  weight in the ground state wave function, we designed an importance sampling that takes into account the weights in all of the one particle-hole excitations. This allows to calculate all block matrices $\tilde{\mathcal{H}}^{\mathbf{k}}$ and $\mathcal{O}^{\mathbf{k}}$ for every wave vector $\mathbf{k}$ in a \emph{single} Monte Carlo simulation.
We tested the method on the example of the SU(3) Heisenberg chain and the Haldane-Shastry model.

In Secs.~\ref{sec:static}-\ref{sec:EGS} we considered properties which can be calculated from the Gutzwiller projected Fermi sea used as the ground state. We reproduced the structure factor $S^{33}(k)$ by a standard VMC and confirmed that the exponent of the singularity at momentum $2 \pi/3$ is the expected $\eta = 4/3$. Next, we derived expressions for the single-mode approximation of the SU(N) Heisenberg models and calculated the corresponding dynamical structure factor. The long-wavelength limit provided the velocity of excitations. We recovered the exact velocity for the Haldane-Shastry model, while for the Heisenberg model, it was about 16\% larger than the value known from the Bethe-Ansatz solution. We got a better approximation using the dynamical VMC, which gave a velocity only about 4\% larger than the exact value. The finite-size scaling of the ground state energy was consistent with a central charge $c=2$.

In Sec.~\ref{sec:Skw} we applied the dynamical VMC to calculate the dynamical structure factor using one particle-hole excitations, up to $L=84$ sites. We compared the $L=18$ site result with the one from exact diagonalization, and the precision at low energies was excellent. For larger system sizes, the support of the $S^{33}(k,\omega)$ follows the two-soliton continuum of the Bethe-Ansatz. Also, the overall weight distribution agrees with the matrix-product-state calculation presented in \cite{DMRG_Binder_PRB2020}. However, a detailed examination reveals that the one particle-hole excitations fail to reproduce the four-soliton excitations. Finally, we analyzed the critical properties: the finite-size scaling of the gap at $k=2\pi/3$ and the power-law behavior of the dynamical structure factor at low energies. In both cases, the behavior followed the expected one from the conformal theory. We also calculated $S^{33}(k,\omega)$ for the Haldane-Shastry model for small ($L=9$, 12, and 15) systems. The method gave the exact weight and energy of the peak at the bottom of the conformal towers and for the lower edge of the continuum between $k=0$ and $2\pi/3$ (the ``des Cloizeaux-Pearson branch'' for the $S=1/2$ Heisenberg model). We identified a class of Gutzwiller projected one particle-hole excitations of the Fermi sea that are exact eigenstates of the Haldane-Shastry model. The only case where the dynamical VMC performed poorly was the scaling of the weight of the bottom of the conformal tower with system size, where it did not seem to give the precise $\eta-1=-1/3$ exponent, but it was closer to $-0.25$ -- the precise origin of the discrepancy is not clear to us.

\begin{acknowledgments}

We thank Ferenc Woynarovich for help on interpretation of the Bethe Ansatz results and Fr\'ed\'eric Mila and Mithilesh Nayak for discussion on the dynamical structure factor and for sharing with us their unpublished results on the dynamical structure factor. D.V. thanks Tamás Molnár for his help with C++ and Linux related issues, and Zoltán Vörös (from the Space Research Institute in Graz) for useful discussions regarding error estimation.

This work was supported by the Hungarian NKFIH Grant No. K 124176 and the BME - Nanonotechnology and Materials Science FIKP grant of EMMI (BME FIKP-NAT).

\end{acknowledgments}

\appendix

%%%%%%%%%%%%%%%%%%%%%%%%%%%%%%%%%%%%%%%%%%%%%%%%%%%%%%%%%%%%%%%%%%
\section{Oscillator strength in SU(N) Heisenberg model}
\label{sec:SUN_double_commutator}
%%%%%%%%%%%%%%%%%%%%%%%%%%%%%%%%%%%%%%%%%%%%%%%%%%%%%%%

In this section we work out the formula of the oscillator strength for the SU(N) spins, given by the double commutator  
\begin{equation}
    f(\mathbf{k}) = \frac{1}{2} \left\langle \left[ \left[ T_{-\mathbf{k}}^a, H \right], T_{\mathbf{k}}^a \right]  \right\rangle,
    \label{eq:fqdc}
\end{equation}
where there is no summation for $a$. The Hamiltonian is 
\begin{equation}
    \mathcal{H} = \sum_{\langle l, l' \rangle} J_{l,l'} \mathbf{T}_{l}\cdot \mathbf{T}_{l'} \;,
\end{equation}
and the operator $T_{\mathbf{k}}^a$ is defined as 
\begin{equation}
    T_{\mathbf{k}}^a \equiv \frac{1}{\sqrt{L}} \sum_{j} e^{-i\mathbf{k}\cdot \mathbf{R}_{j}} T_{j}^a \;.
\end{equation}
For generality we consider a model in arbitrary spatial dimension, the  
$\mathbf{R}_{j}$ denotes the position of $j^{\text{th}}$ site.
Inserting the expressions above into the  double commutator in Eq.~(\ref{eq:fqdc}), the oscillator strength becomes:
%\begin{equation}
%    f(q) = \frac{1}{2} \sum_{\langle l, l' \rangle} J_{l,l'} \left\langle \left[ \left[ T_{-q}^a, \mathbf{T}_{l}\cdot \mathbf{T}_{l'} \right], T_{q}^a \right]  \right\rangle.
%\end{equation}
%
%Putting these in the expression of the oscillator strength:
\begin{equation}
    f(\mathbf{k}) = \frac{1}{2L}  \sum_{\langle l, l' \rangle} J_{l,l'} \sum_{j,j'} e^{i\mathbf{k}\cdot (\mathbf{R}_{j'} - \mathbf{R}_{j}) } \left\langle \left[ \left[ T_{j'}^a, \mathbf{T}_{l}\cdot \mathbf{T}_{l'} \right], T_{j}^a \right]  \right\rangle.
    \label{eq:fqApp}
\end{equation}
%The operators $T_{j}^a$ on different sites commute, so if $i\neq j$ the commutator $\left[ T_{i}^a, T_{j}^b \right] = 0$ for $\forall a, b$. This implies that if $j' \neq l $ and $j' \neq l'$, than $\left[ T_{j'}^a, \mathbf{T}_{l}\cdot \mathbf{T}_{l'} \right] = 0$, and if $j \neq l $ and $j \neq l'$ than independently of the value of $j'$ the commutator $\left[ \left[ T_{j'}^a, \mathbf{T}_{l}\cdot \mathbf{T}_{l'} \right], T_{j}^a \right] = 0$. The only non-zero terms are those for which both $j$ and $j'$ takes one of the values of $l$ and $l'$. 
%\begin{align}
%    f(q) =\\ 
%    \frac{1}{2L}  \sum_{\langle l, l' \rangle} J_{l,l'} \big( & \underbrace{e^{i\mathbf{k}\cdot (\mathbf{R}_{l\ } - \mathbf{R}_{l\ }) } }_{1}\left\langle \left[ \left[ T_{l}^a, \mathbf{T}_{l}\cdot \mathbf{T}_{l'} \right], T_{l}^a \right]  \right\rangle \nonumber\\
%    + &\underbrace{e^{i\mathbf{k}\cdot (\mathbf{R}_{l'} - \mathbf{R}_{l'}) }}_{1} \left\langle \left[ \left[ T_{l'}^a, \mathbf{T}_{l}\cdot \mathbf{T}_{l'} \right], T_{l'}^a \right]  \right\rangle \nonumber\\
%    + &e^{i\mathbf{k}\cdot (\mathbf{R}_{l'} - \mathbf{R}_{l\ }) } \left\langle \left[ \left[ T_{l'}^a, \mathbf{T}_{l}\cdot \mathbf{T}_{l'} \right], T_{l}^a \right]  \right\rangle \nonumber\\
%    + &e^{i\mathbf{k}\cdot (\mathbf{R}_{l\ } - \mathbf{R}_{l'}) } \left\langle \left[ \left[ T_{l}^a, \mathbf{T}_{l}\cdot \mathbf{T}_{l'} \right], T_{l'}^a \right]  \right\rangle \big)\nonumber
%\end{align}
 Since the operators $T_{j}^a$ on different sites commute, the only non-zero terms are those for which both $j$ and $j'$ takes one of the values of $l$ and $l'$:
 \begin{align}
    f(\mathbf{k})  
     =\frac{1}{2L}  \sum_{\langle l, l' \rangle} & J_{l,l'} \big(
      \left\langle \left[ \left[ T_{l}^a, \mathbf{T}_{l}\cdot \mathbf{T}_{l'} \right], T_{l}^a \right]  \right\rangle 
   \nonumber\\ &
    + \left\langle \left[ \left[ T_{l'}^a, \mathbf{T}_{l}\cdot \mathbf{T}_{l'} \right], T_{l'}^a \right]  \right\rangle 
    \nonumber\\&
    + e^{i\mathbf{k}\cdot (\mathbf{R}_{l'} - \mathbf{R}_{l\ }) } \left\langle \left[ \left[ T_{l'}^a, \mathbf{T}_{l}\cdot \mathbf{T}_{l'} \right], T_{l}^a \right]  \right\rangle 
    \nonumber\\&
    + e^{i\mathbf{k}\cdot (\mathbf{R}_{l\ } - \mathbf{R}_{l'}) } \left\langle \left[ \left[ T_{l}^a, \mathbf{T}_{l}\cdot \mathbf{T}_{l'} \right], T_{l'}^a \right]  \right\rangle \big) \;.
    \label{eq:doublcomm_temp}
\end{align}

Let us calculate the double commutators of the SU(3) invariant quantity $\sum_a [[T^a_l,\mathbf{T}_l \cdot \mathbf{T}_{l'}],T^a_l]$ with the help of the commutation relations (\ref{eq:su3_com}) of the su($N$) algebra: 
\begin{align}
\sum_a [[T^a_l,\mathbf{T}_l \cdot \mathbf{T}_{l'}],T^a_l] 
&= \sum_{a,b} [[T^a_l,T^b_l T^b_{l'}],T^a_l] \nonumber\\
&=  \sum_{a,b} [[T^a_l,T^b_l],T^a_l] T^b_{l'} \nonumber\\
&=  i \sum_{a,b,c} f_{abc} [T^c_l,T^a_l] T^b_{l'} \nonumber\\
&=  - \sum_{a,b,c,d} f_{abc} f_{cad} T^d_l T^b_{l'} \nonumber\\
&=  - \sum_{b,d} N \delta_{bd} T^d_l T^b_{l'} \nonumber\\
&= - N  \mathbf{T}_l \cdot \mathbf{T}_{l'} 
\end{align}
where we used that $\sum_{a,c} f_{abc} f_{cad} = N \delta_{bd}$ \cite{SUrels_SciPostPhysLectNotes.21}.
Since all of the terms in the sum contribute equally, we may write
\begin{equation}
[[T^a_l,\mathbf{T}_l \cdot \mathbf{T}_{l'}],T^a_l] = - \frac{N}{N^2-1}  \mathbf{T}_l \cdot \mathbf{T}_{l'} \;.
\end{equation}
Similar considerations apply to the case when the $T^a$ operators are on different sites: 
\begin{equation}
[[T^a_l,\mathbf{T}_l \cdot \mathbf{T}_{l'}],T^a_{l'}] =  \frac{N}{N^2-1}  \mathbf{T}_l \cdot \mathbf{T}_{l'}  \;.
\end{equation}
Inserting the equations above into the expression (\ref{eq:doublcomm_temp}) of the oscillator strength, we get\begin{equation}
    f(\mathbf{k}) =
     \frac{N}{L(N^2-1)} \sum_{\langle l, l' \rangle} J_{l,l'} \left( \cos \mathbf{k} \cdot \mathbf{d}_{l'l} -1 \right) \left\langle \mathbf{T}_l \cdot \mathbf{T}_{l'} \right\rangle \;,
\end{equation}
where $\mathbf{d}_{l',l} = \mathbf{R}_{l'} - \mathbf{R}_l $.
For a translationally invariant one-dimensional model this simplifies to
\begin{equation}
    f(k) =
     - \frac{N}{(N^2-1)} \sum_{\langle l \rangle} J_{l} \sin^2 \frac{k l}{2}  \left\langle \mathbf{T}_0 \cdot \mathbf{T}_{l} \right\rangle \;.
     \label{eq:f(q)1D}
\end{equation}

% Phys. Rev. {\bf 91}, 1291, 1301 (1953); {\bf 94}, 262 (1954); R. P. Feynman and M. Cohen ibid. {\bf 102}, 1189 (1956).
%
%S. M. Girvin, A. H. MacDonald, and P. M. Platzman,
%{\it Magneto-roton theory of collective excitations in the fractional quantum Hall effect}, 
%\href{https://doi.org/10.1103/PhytsRevB.33.2481}{Phys. Rev. B {\bf{33}}, 2481 (1985)}
%

%%%%%%%%%%%%%%%%%%%%%%%%%%%%%%%%%%%%%%%%%%%%%%%%%%%%%%%
\section{The generalized eigenvalue problem}
\label{sec:gen_eigenvalue}
%%%%%%%%%%%%%%%%%%%%%%%%%%%%%%%%%%%%%%%%%%%%%%%%%%%%%%%

Not all of the states (\ref{particle-hole excited states}) or (\ref{excited states of Becca}) are linearly independent. The linear dependencies show up as zero eigenvalues of the overlap matrix.. In order to solve the generalized eigenvalue problem, the overlap matrix has to be positive definit, therefore we have to perform a basis transformation to remove the numerically zero eigenvalues. This could be problematic if some of the positive eigenvalues of the overlap matrix were close to the numerical error of the zero eigenvalues, and so we could not distinguish between positive eigenvalues and zero eigenvalues. Fortunately, the eigenvalues of the overlap matrix have a gap of many orders of magnitude, which well separates  the positive eigenvalues from the numerically zero eigenvalues. If we perform the basis transformation $| \mathbf{k}, \mathbf{R}, \sigma \rangle \rightarrow   \frac{1}{\sqrt{2}} \left( | \mathbf{k}, \mathbf{R}, \mathsf{A} \rangle - | \mathbf{k}, \mathbf{R}, \mathsf{B} \rangle \right) $ (the relevant excited states for the measurement of $T^3$), then the number of positive eigenvalues of the overlap matrix increases linearly with the relative momentum $q$ as $\frac{L}{2\pi} q$ in the interval $q \in \lbrace 0, \frac{2 \pi}{3}\rbrace$, then it saturates for $\frac{2 \pi}{3} \leq q  \leq \frac{4 \pi }{3}$, and it decreases linearly again until reaching 0 at $q=2\pi$. The number of the linearly independent states for a given momentum $q$ is in fact equal to the number of possible one particle-hole excitations in the Fermi sea of the same momentum $q$.

In order to find the eigenstates (\ref{eigenstates of the projected Hamiltonian}) of a block matrix of the projected Hamiltonian $\tilde{\mathcal{H}}^{\mathbf{k}}$, we must first diagonalize the corresponding block of the overlap matrix as
\begin{equation}
    \mathcal{\overline{O}^{\mathbf{k}}} = \mathcal{U}^{\dagger \mathbf{k}} \mathcal{O^{\mathbf{k}}} \mathcal{U^{\mathbf{k}}},
\end{equation}
where $\mathcal{\overline{O}^{\mathbf{k}}}$ is a     diagonal matrix containing the sorted eigenvalues of $\mathcal{O}^{\mathbf{k}}$, and $\mathcal{U^{\mathbf{k}}}$ is the matrix having the eigenstates of $\mathcal{O}^{\mathbf{k}}$ in its columns, in the order of the corresponding eigenvalues in $\mathcal{\overline{O}^{\mathbf{k}}}$. Next we calculate $\mathcal{\overline{H}^{\mathbf{k}}} \equiv \mathcal{U}^{\dagger \mathbf{k}} \tilde{\mathcal{H}}^{\mathbf{k}} \mathcal{U^{\mathbf{k}}}$ and solve the generalized eigenvalue problem for the blocks of the block matrices $\mathcal{\overline{O}^{\mathbf{k}}}$ and $\mathcal{\overline{H}^{\mathbf{k}}}$, which correspond to the subspace of eigenstates of $\mathcal{O^{\mathbf{k}}}$ having positive eigenvalues. The dimension of the blocks of the block matrices $\mathcal{\overline{O}^{\mathbf{k}}}$ and $\mathcal{\overline{H}^{\mathbf{k}}}$ is equal to the number of positive eigenvalues of $\mathcal{\overline{O}^{\mathbf{k}}}$, therefore the dimension of the eigenstates of $\mathcal{\overline{H}^{\mathbf{k}}}$ (obtained from the generalized eigenvalue problem) is also the number of the positive eigenvalues. In order to obtain the eigenstates of $\mathcal{\overline{H}^{\mathbf{k}}}$ in the original basis of particle-hole excitations, we have to put in these eigenstates zeros for each zero eigenvalue of $\mathcal{\overline{O}^{\mathbf{k}}}$, and then we can transform them back by acting with $U^{ \mathbf{k}}$. This way we arrive to the eigenstates (\ref{eigenstates of the projected Hamiltonian}) of the projected Hamiltonian  we were looking for, in the basis of the states (\ref{excited states of Becca}). These are the eigenstates used in Eq.~(\ref{sandwich of T3 expressed with the overlap matrix}).

%%%%%%%%%%%%%%%%%%%%%%%%%%%%%%%%%%%%%%%%%%%%%%%%%%%%%%%
\section{Monte Carlo evaluation of the matrices $\tilde{\mathcal{H}}$ and $\mathcal{O}$ using importance sampling}
\label{sec:Monte_Carlo_of_H_and_O}
%%%%%%%%%%%%%%%%%%%%%%%%%%%%%%%%%%%%%%%%%%%%%%%%%%%%%%%

In order to evaluate the matrices $\tilde{\mathcal{H}}^{\mathbf{k}}_{\mathbf{R}, \sigma;\mathbf{R'},\sigma'}$ and $\mathcal{O}^{\mathbf{k}}_{\mathbf{R}, \sigma;\mathbf{R'},\sigma'}$ it is useful to insert the identity operator $\mathbf{I} = \sum_x |x\rangle \langle x|$ into Eqs.~(\ref{eq:HRR}) and (\ref{eq:ORR})
\begin{subequations}
\label{eq:HOx}
\begin{align}
    &\tilde{\mathcal{H}}^{\mathbf{k}}_{\mathbf{R}, \sigma;\mathbf{R'},\sigma'} = \sum_x  \langle \mathbf{k} , \mathbf{R}, \sigma | x\rangle \langle x|  \mathcal{H} | \mathbf{k} , \mathbf{R'}, \sigma' \rangle 
    \\
    &\mathcal{O}^{\mathbf{k}}_{\mathbf{R}, \sigma;\mathbf{R'},\sigma'} =\sum_x  \langle \mathbf{k} , \mathbf{R}, \sigma | x\rangle \langle x| \mathbf{k} , \mathbf{R'}, \sigma' \rangle ,
\end{align}
\end{subequations}
where the orthonormal basis set of states $\lbrace | x \rangle \rbrace$ corresponds to real space configurations of particles having the same number of particles of each color as $|\mathbf{k} , \mathbf{R}, \sigma \rangle$.
%, since for zero external field the Fermi sea has equal number of particles of each color, and $|\mathbf{k} , \mathbf{R}, \sigma \rangle $ does not change the numbers of particles of given colors. 
For small system sizes (we did it until $L=21$, see Tab.~\ref{tab:ED_vari} for a comparison with ED for some selected quantities), the expressions above can be evaluated directly by going through each configuration $|x \rangle$ of the Hilbert space and calculating $\langle x | \mathcal{H} | \mathbf{k} , \mathbf{R}, \sigma \rangle $ and $\langle x | \mathbf{k} , \mathbf{R}, \sigma \rangle $. In this manner we get numerically exact values for the matrices $\tilde{\mathcal{H}}^{\mathbf{k}}$ and $\mathcal{O}^{\mathbf{k}}$, and solving the generalized eigenvalue equation, we get the excited states and the dynamical structure factor.

 However, for larger system sizes the direct evaluation becomes difficult, as the size of the Hilbert space grows exponentially. Instead, one can use a Monte Carlo method to evaluate the Hamiltonian and overlap matrix Eqs.~(\ref{eq:HOx}) by random sampling the states $| x \rangle$. This is rather inefficient unless the sampling takes into account the weight of the configuration $| x \rangle$.  This can be achieved by importance sampling. To evaluate a sum by importance sampling one rewrites the sum as $\sum_x f(x) = \sum_x g(x) P(x)$, where $\sum_x P(x) = 1$, $P(x) \geq 0$ $\forall x$ and $g(x) = f(x) / P(x)$. The configurations $| x \rangle$ are sampled based on the probability distribution $P(x)$, and for each sampled configuration $| x \rangle$ we measure $g(x)$. $g(x)$ might diverge for configurations which have $P(x) = 0$, but these configurations are not reached by importance sampling. Therefore, it is preferable to choose a $P(x)$ which is non-zero for each configuration $x$ for which $f(x)$ is non-zero.
 In this spirit, we modify the Eqs.~(\ref{eq:HOx}) by multiplying and dividing by $P(x)$:
\begin{subequations}
\label{imporatnce sampling of H and O by a general probability distribution}
\begin{align}
    \tilde{\mathcal{H}}^{\mathbf{k}}_{\mathbf{R} \sigma,\mathbf{R'},\sigma'} &= 
    \sum_x  \frac{ \langle \mathbf{k} , \mathbf{R}, \sigma | x\rangle }{ \sqrt{P(x)} } 
    \frac{ \langle x|  \mathcal{H} | \mathbf{k} , \mathbf{R'}, \sigma' \rangle }{ \sqrt{P(x)} }  P(x) \;,
    \\
    \mathcal{O}^{\mathbf{k}}_{\mathbf{R} \sigma,\mathbf{R'},\sigma'} &= 
    \sum_x  \frac{ \langle \mathbf{k} , \mathbf{R}, \sigma | x\rangle }{ \sqrt{P(x)} } 
    \frac{ \langle x | \mathbf{k} , \mathbf{R'}, \sigma' \rangle }{ \sqrt{P(x)} } 
    P(x) .
\end{align}
\end{subequations}

The probability distribution $P(x)$ can be chosen in many ways, here we give a brief overview of the choices used in previous papers. Li and Yang chose the probability distribution
\begin{equation}
\label{original paper probability distribution}
 P^{ \mathbf{k}  }(x) = \frac{ \sum_{ \mathbf{q},\sigma} | \langle x |  \mathbf{k} , \mathbf{q}, \sigma  \rangle |^2 }{ \sum_x \sum_{ \mathbf{q},\sigma} | \langle x |  \mathbf{k} , \mathbf{q}, \sigma  \rangle |^2 },
\end{equation}
also followed by \cite{Excited_states_above_Fermi_sea}. This probability distribution was used to sample the block matrices $\tilde{\mathcal{H}}^{\mathbf{k}}$ and $\mathcal{O}^{\mathbf{k}}$, which meant a separate Monte Carlo simulation for each $\mathbf{k}$ \cite{First_S_q_w_latter,*Yang_Li_2011PhRvB..83f4524Y}.

On the other hand, Ferrari {\it et al.} sampled according to the weight of $|x \rangle$ in the ground state \cite{Becca},
\begin{equation}
\label{Beccas probability distribution}
P(x) = \frac{| \langle x |  P_{\text{G} } | \text{FS} \rangle |^2 }{ \sum_x | \langle x |  P_{\text{G} } | \text{FS} \rangle |^2 },
\end{equation}
where $\langle x |  P_{\text{G} } | \text{FS} \rangle$ is a product of real Slater determinants (\ref{eq:real_space_FS}). The advantage is the ability to sample all the block matrices $\tilde{\mathcal{H}}^{\mathbf{k}}$ and $\mathcal{O}^{\mathbf{k}}$ simultaneously. Furthermore, the terms 
\begin{equation}
    \frac{ \langle x|  \mathcal{H} | \mathbf{k} , \mathbf{R}, \sigma \rangle }{ \langle x |  P_{\text{G} } | \text{FS} \rangle  } =
    \frac{1}{\sqrt{L}} \sum_{\mathbf{R'}} e^{i\mathbf{k} \cdot \mathbf{R'}} \frac{ \langle x| \mathcal{H} P_{\text{G}} f^{{\dagger}}_{\mathbf{R+R'},\sigma} f^{\phantom{\dagger}}_{\mathbf{R'},\sigma} |\text{FS} \rangle}{\langle x |  P_{\text{G} } | \text{FS} \rangle}
\end{equation}
and 
\begin{equation}
    \frac{ \langle x| \mathbf{k} , \mathbf{R}, \sigma \rangle }{ \langle x |  P_{\text{G} } | \text{FS} \rangle   } = 
    \frac{1}{\sqrt{L}} \sum_{\mathbf{R'}} e^{i\mathbf{k} \cdot \mathbf{R'}} \frac{ \langle x| P_{\text{G}} f^{{\dagger}}_{\mathbf{R+R'},\sigma} f^{\phantom{\dagger}}_{\mathbf{R'},\sigma} |\text{FS} \rangle}{\langle x |  P_{\text{G} } | \text{FS} \rangle} 
    \;,  
\end{equation}
appearing in the expressions (\ref{imporatnce sampling of H and O by a general probability distribution}) can be calculated very efficiently using the rank-1 determinant update, since they reduce to quotients of real Slater determinants (\ref{eq:slater_determinant}) which differ in a single column only.
However, configurations which are important for the excited states, but unimportant for the ground state will be sampled rarely: the $ |\langle x_1 |  P_{\text{G} } | \text{FS} \rangle | \gg |\langle x_2 |  P_{\text{G} } | \text{FS}  \rangle|$ condition does not imply $|\langle x_1| \mathbf{k} , \mathbf{R}, \sigma \rangle| \gg |\langle x_2| \mathbf{k} , \mathbf{R}, \sigma \rangle|$ 
nor 
$| \langle x_1|  \mathcal{H} | \mathbf{k} , \mathbf{R}, \sigma \rangle | \gg | \langle x_2|  \mathcal{H} | \mathbf{k} , \mathbf{R}, \sigma \rangle |$. Thus, the $|x_2 \rangle$ may be just as important for some excited states as $|x_1\rangle$ is for the ground state, and still it will be sampled with much smaller probability.

Mei and Wen used an importance sampling similar to (\ref{Beccas probability distribution}), with the difference of working in the subspace of $S^z_{T} = 1$, and replacing the $P_{\text{G}}|\text{FS}\rangle $ with the lowest mean field particle-hole state in this subspace \cite{Mei_Wen_arxiv_2015}.

Extending the sum over $\mathbf{q}$ and $\sigma$ to a sum including all $\mathbf{k}$-s in the probability distribution (\ref{original paper probability distribution}) 
\begin{equation}
\label{original paper probability distribution extended}
P(x) = \frac{ \sum_{ \mathbf{k}, \mathbf{q},\sigma} | \langle x |  \mathbf{k} , \mathbf{q}, \sigma  \rangle |^2 }{ \sum_x \sum_{ \mathbf{k}, \mathbf{q},\sigma} | \langle x |  \mathbf{k} , \mathbf{q}, \sigma  \rangle |^2 }.
\end{equation}
would make it possible to sample all the block matrices $\tilde{\mathcal{H}}^{\mathbf{k}}$ and $\mathcal{O}^{\mathbf{k}}$ simultaneously. However, while the weights $\langle x |  P_{\text{G} } | \text{FS} \rangle$ in (\ref{Beccas probability distribution}) are real, the weights $\langle x | \mathbf{k}, \mathbf{q}, \sigma  \rangle $ are products of Slater determinants out of which at least one is complex. The reason is, that the Slater determinant (\ref{eq:slater_determinant}) of color $\sigma$ contains the one-particle eigenstates of wave vectors $\mathbf{q}$ and $\mathbf{k} + \mathbf{q}$, but not their pairs with wave vectors $-\mathbf{q}$ and $-\mathbf{k} - \mathbf{q}$ (\ref{fig:particle-hole excitation}), so that no basis transformation can be done to make these states real, as explained at the end of section (\ref{sec:PGFS}). 

In order to work with real Slater-determinants, we used the probability distribution
\begin{equation}
\label{max of abs of g matrix}
    P(x) = \frac{ \max_{\mathbf{R},\mathbf{R'},\sigma} | \langle x |\mathbf{R},\mathbf{R'},\sigma \rangle | }{ \sum_{x} \max_{\mathbf{R},\mathbf{R'},\sigma} | \langle x |\mathbf{R},\mathbf{R'},\sigma \rangle | },
\end{equation}
where we introduced the notation
\begin{equation}
\label{g_matrix}
|\mathbf{R},\mathbf{R'},\sigma \rangle \equiv P_{\text{G}} f^{{\dagger}}_{\mathbf{R},\sigma} f^{\phantom{\dagger}}_{\mathbf{R'},\sigma} |\text{FS} \rangle.
\end{equation}
The weights of this probability distribution are real, since
\begin{equation}
    \langle x |\mathbf{R},\mathbf{R'},\sigma \rangle = \langle x | P_{\text{G}} f^{{\dagger}}_{\mathbf{R},\sigma} f^{\phantom{\dagger}}_{\mathbf{R'},\sigma} |\text{FS} \rangle = \langle x' | \text{FS} \rangle,
\end{equation}
is a product of real Slater determinants (\ref{eq:real_space_FS}), which is the weight of the configuration $|x' \rangle \equiv f^{{\dagger}}_{\mathbf{R'},\sigma} f^{\phantom{\dagger}}_{\mathbf{R},\sigma} P_{\text{G}}|x \rangle $.

With this notation the definition of the states $|\mathbf{k},\mathbf{R},\sigma \rangle$ from Eq.~(\ref{excited states of Becca}) can be rewritten as
\begin{equation}
\label{fourier transform of g_matrix}
|\mathbf{k},\mathbf{R},\sigma \rangle = \frac{1}{\sqrt{L}} \sum_{\mathbf{R'}} e^{i\mathbf{k} \cdot \mathbf{R'}} |\mathbf{R} + \mathbf{R'},\mathbf{R'},\sigma \rangle.
\end{equation}
Comparing this with Eq.~(\ref{relation between Dalla and Becca states}) the states $|\mathbf{R},\mathbf{R'},\sigma \rangle$ correspond to Fourier transforming the particle-hole excitations $|\mathbf{k},\mathbf{q},\sigma \rangle$ in both $\mathbf{k}$ and $\mathbf{q}$. In Eq.~(\ref{max of abs of g matrix}) we summed over every index of the states $|\mathbf{R},\mathbf{R'},\sigma \rangle $, therefore using the probability distribution (\ref{max of abs of g matrix}) as a guiding function we are sampling each block matrix $\tilde{\mathcal{H}}^{\mathbf{k}}$ and $\mathcal{O}^{\mathbf{k}}$ simultaneously.
The choice of the maximum norm in (\ref{max of abs of g matrix}) is arbitrary, in fact, any norm of $ \langle x |\mathbf{R},\mathbf{R'},\sigma \rangle $ is suitable for importance sampling. The norm in Eq.~(\ref{max of abs of g matrix}) is a special case of the $p$-norm  
\begin{equation}
 P(x) =  \frac{ \left( \sum_{|  \mathbf{R} , \mathbf{R'}, \sigma \rangle} | \langle x |\mathbf{R},\mathbf{R'},\sigma \rangle |^p \right)^{1/p} }{\sum_x  \left( \sum_{|  \mathbf{R} , \mathbf{R'}, \sigma \rangle} | \langle x |\mathbf{R},\mathbf{R'},\sigma \rangle |^p \right)^{1/p} },
\end{equation} 
with $p = \infty $. Using a norm of $ \langle x |\mathbf{R},\mathbf{R'},\sigma \rangle $ is useful, because if this norm is small (large), than based on Eq.~(\ref{fourier transform of g_matrix}) the norm of $\langle x | \mathbf{k},\mathbf{R},\sigma \rangle$ will be small (large) as well, and the latter is present in both $\tilde{\mathcal{H}}^{\mathbf{k}}_{\mathbf{R} \sigma,\mathbf{R'},\sigma'}$ and $\mathcal{O}^{\mathbf{k}}_{\mathbf{R} \sigma,\mathbf{R'},\sigma'}$ as can be seen from Eqs.~(\ref{imporatnce sampling of H and O by a general probability distribution}).

On the one hand, this importance sampling is slower than that of Eq.~(\ref{Beccas probability distribution}) used by Ferrari {\it et al.} in Ref.~\cite{Becca}, since in each elementary step we have to calculate the $N L^2$ elements of $\langle x |\mathbf{R},\mathbf{R'},\sigma \rangle$. But these elements are products of Slater determinants out of which one differs from those in  $ \langle x | P_{\text{G}} | \text{FS} \rangle$ in a single column only, so they can be calculated efficiently with a rank-1 determinant update. On the other hand, the configurations which are important for the excited states only are sampled with higher probabilities, thus yielding a better statistics for the block matrices $\tilde{\mathcal{H}}^{\mathbf{k}}$ and $\mathcal{O}^{\mathbf{k}}$ with $\mathbf{k} \neq \mathbf{0}$.

The numerator $\sum_x  \max_{  \mathbf{R'} , \mathbf{R''}, \sigma } | \langle x |  \mathbf{R'} , \mathbf{R''}, \sigma  \rangle |$ 
of the probability distribution (\ref{max of abs of g matrix}) is independent of the configuration $|x\rangle$, it multiplies both the $\tilde{\mathcal{H}}^{\mathbf{k}}$ and the $\mathcal{O}^{\mathbf{k}}$. Consequently, it falls out from the generalized eigenvalue problem (\ref{generalized eigenvalue problem}), and we do not have to measure it at all. Thus, the measurement of $\tilde{\mathcal{H}}$ and $\mathcal{O}$ for a given configuration $|x\rangle$ consists of calculating the quantities

\begin{equation}
    \frac{ \langle x|  \mathcal{H} | \mathbf{k} , \mathbf{R}, \sigma \rangle }{ \sqrt{\max_{   \mathbf{R'} , \mathbf{R''}, \sigma }| \langle x |  \mathbf{R'} , \mathbf{R''}, \sigma  \rangle |} }
\end{equation}
and
\begin{equation}
    \frac{ \langle x | \mathbf{k} , \mathbf{R}, \sigma \rangle }{ \sqrt{\max_{   \mathbf{R'} , \mathbf{R''}, \sigma }| \langle x |  \mathbf{R'} , \mathbf{R''}, \sigma  \rangle |} }.
\end{equation}
The difficulty is in measuring $\langle x|  \mathcal{H} | \mathbf{k} , \mathbf{R}, \sigma \rangle$, since $\langle x| \mathbf{k} , \mathbf{R}, \sigma \rangle$ can be calculated from $\langle x|\mathbf{R},\mathbf{R'},\sigma \rangle$ using Eq.~(\ref{fourier transform of g_matrix}), and $\langle x |\mathbf{R},\mathbf{R'},\sigma \rangle$ was already calculated during importance sampling.

Finally, we employed the Metropolis–Hastings algorithm for the sampling of the configurations. In each elementary step we randomly choose two sites having particles of different colors with uniform probability, and we exchange them by the acceptance probability 
\begin{equation}
A(x \rightarrow x') =
  \begin{cases}
\displaystyle{\frac{P(x')}{P(x)}}, & \text{if}\ P(x') < P(x), \\
 1, &\text{if}\ P(x') > P(x),
  \end{cases}
\end{equation}
 where $|x'\rangle$ is the configuration resulting from $|x\rangle$ after exchanging the two particles at the chosen sites.

In order to get independent measurements, they should be separated by a number of elementary steps which is greater than the correlation time. We estimated the correlation time by measuring how many elementary steps are needed after equilibration to get $L$ accepted elementary steps, where $L$ is the number of lattice sites. Since $L$ pair exchanges are enough to reach any configuration from the present configuration (Fisher-Yates shuffles), we assume that after $L$ accepted pair exchanges the configuration is not correlated with the previous one.

%%%%%%%%%%%%%%%%%%%%%%%%%%%%%%%%%%%%%%%%%%%%%%%%%%%%%%%
\section{Estimation of statistical errors }
\label{sec:error estimation}
%%%%%%%%%%%%%%%%%%%%%%%%%%%%%%%%%%%%%%%%%%%%%%%%%%%%%%%

We run the program typically a hundred times for each system size. Let us denote the number of runs by $M(L)$ for a system with $L$ sites. For the structure factor, each run included about $5 \times 10^{6}$, while for the dynamical structure factor $5-10 \times 10^5$ measurements, separated by elementary steps which number corresponds to the correlation time. 
In the $i^{\text{th}}$ run we obtained the  average of measurements $Q_i(L)$, $i=1,\dots,M(L)$. The average 
\begin{equation}
  \bar Q(L) = \frac{1}{M(L)} \sum_{i=1}^{M(L)} Q_i(L)
\end{equation}
is the result of the MC calculation, with the standard error
\begin{equation}
    \sigma_{Q}(L) = \sqrt{\frac{\sum_{i=1}^{M(L)}
    [Q_i(L) -\bar Q(L)]^2}{M(L)[ M(L) - 1 ] }}.
\end{equation}
We plot the above standard errors in the figures.

Some quantities were calculated by fitting functions to the data and optimizing the parameters of the functions by the non-linear least squares method, using scipy.curve\_fit. The errors of the optimized parameters were estimated by passing the $\sigma_{Q}(L)$ of the data we wanted to fit on, setting the flag absolute\_sigma = True, and taking the square root of the returned variance.

For the estimation of the error of the central charge we used the error propagation formula
\begin{equation}
    \sigma_{\frac{A}{B}} = \frac{A}{B}  \sqrt{ \left( \frac{\sigma_A}{A} \right)^2 + \left( \frac{\sigma_B}{B} \right)^2 },
\end{equation}
where in our case $A = v c$ and $B = v$.
%https://thefactfactor.com/facts/pure_science/physics/propagation-of-errors/9502/
%https://en.wikipedia.org/wiki/Propagation_of_uncertainty

%\bibliography{su3_chain_PRB} 
\bibliography{su3_corrected_by_Dani.bib} 
\end{document}